\documentclass[sigconf]{acmart}
\usepackage{amsmath,amsfonts}
\usepackage{algorithm,algorithmic}
\usepackage{multirow,multicol}
\usepackage{graphicx}
\usepackage{textcomp}
\usepackage{xcolor}
\usepackage{booktabs}
\usepackage{comment}
\usepackage{footmisc}
\usepackage{mathrsfs}
\usepackage{enumitem}
\usepackage{array}
\usepackage{float}
\usepackage{hyperref}
\usepackage{units}
\usepackage{subfigure} 
\AtBeginDocument{%
	\providecommand\BibTeX{{%
			\normalfont B\kern-0.5em{\scshape i\kern-0.25em b}\kern-0.8em\TeX}}}
		
\begin{document}
\fancyhead{}
\title{Memory-efficient Embedding for Recommendations}

\author{Xiangyu Zhao, Haochen Liu,\\Hui Liu, Jiliang Tang}
\affiliation{
	\institution{Michigan State University}
}
\email{{zhaoxi35,liuhaoc1,liuhui7,tangjili}@msu.edu}

\author{Weiwei Guo, Jun Shi, Sida Wang,\\ Huiji Gao, Bo Long}
\affiliation{
	\institution{Linkedin Corporation}
}
\email{{wguo,jshi,sidwang,hgao,blong}@linkedin.com}

\renewcommand{\shortauthors}{Paper ID: }

\begin{abstract}
Practical large-scale recommender systems usually contain thousands of feature fields from users, items, contextual information, and their interactions. Most of them empirically allocate a unified dimension to all feature fields, which is memory inefficient. Thus it is highly desired to assign different embedding dimensions to different feature fields according to their importance and predictability. Due to the large amounts of feature fields and the nuanced relationship between embedding dimensions with feature distributions and neural network architectures, manually allocating embedding dimensions in practical recommender systems can be very difficult. To this end, we propose an AutoML based framework (AutoDim) in this paper, which can automatically select dimensions for different feature fields in a data-driven fashion. Specifically, we first proposed an end-to-end differentiable framework that can calculate the weights over various dimensions in a soft and continuous manner for feature fields, and an AutoML based optimization algorithm; then we derive a hard and discrete embedding component architecture according to the maximal weights and retrain the whole recommender framework. We conduct extensive experiments on benchmark datasets to validate the effectiveness of the AutoDim framework.  
\end{abstract}

\maketitle
\section{Introduction}
\label{sec:introduction}
With the explosive growth of the world-wide web, huge amounts of data have been generated, which results in the increasingly severe information overload problem, potentially overwhelming users~\cite{chang2006survey}. Recommender systems can mitigate the information overload problem through suggesting personalized items that best match users' preferences~\cite{linden2003amazon,breese1998empirical,mooney2000content,resnick1997recommender,ricci2011introduction,Bao2015Recommendations}. Recent years have witnessed the increased development and popularity of deep learning based recommender systems (DLRSs)~\cite{zhang2017deep,nguyen2017personalized,wu2016personal}, which outperform traditional recommendation techniques, such as collaborative filtering and learning-to-rank, because of their strong capability of feature representation and deep inference~\cite{zhang2019deep}. 

\begin{figure}[t]
	\centering
	\includegraphics[width=\linewidth]{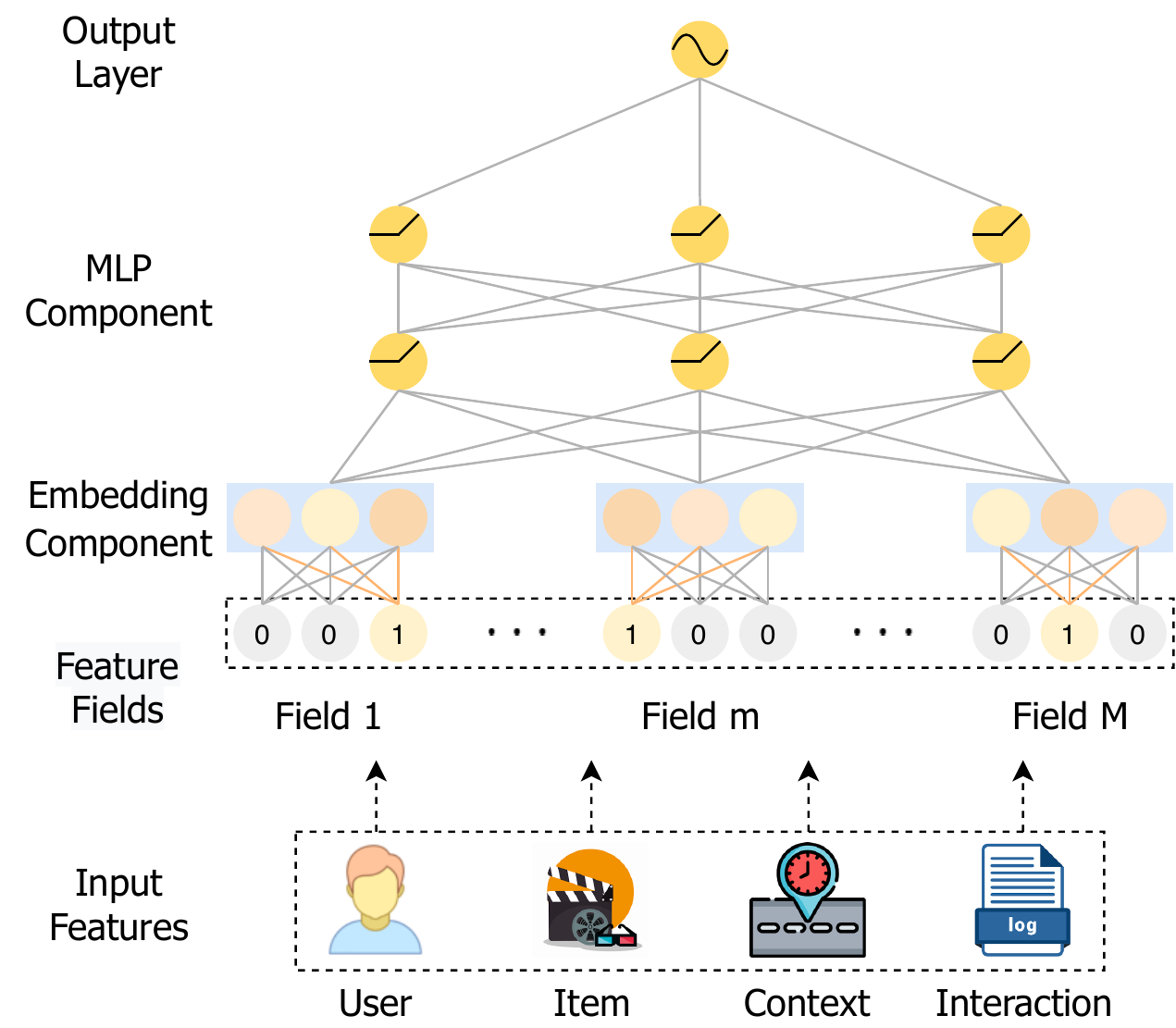}
	\caption{The Typically DLRS architecture.}
	\label{fig:Fig1_DLRS}
\end{figure}

\begin{figure*}
	\centering
	\includegraphics[width=\linewidth]{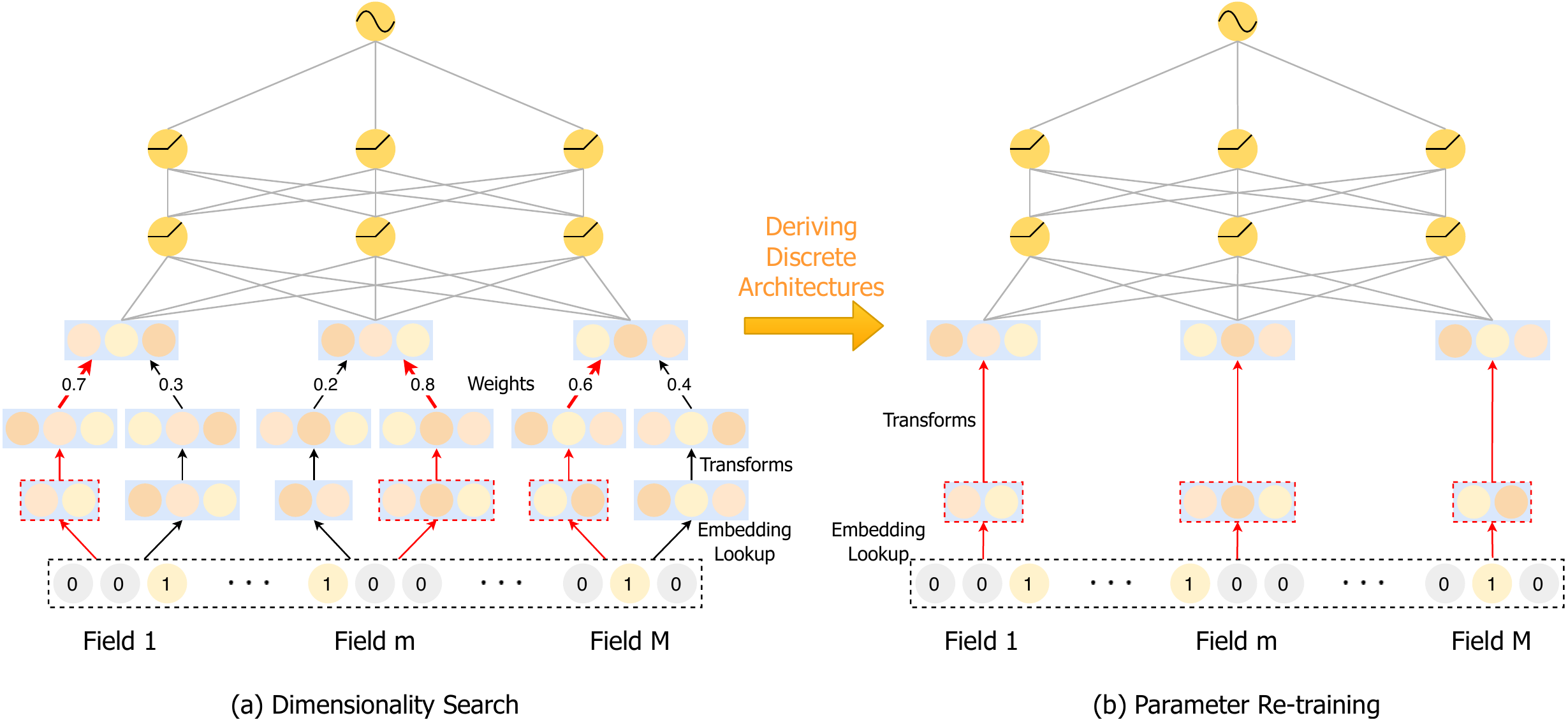}
	\caption{Overview of the proposed framework.}
	\label{fig:Fig2_Framework}
\end{figure*}

Real-world recommender systems typically involve a massive amount of categorical feature fields from users (e.g. occupation and userID), items (e.g. category and itemID), contextual information (e.g. time and location), and their interactions (e.g. user's purchase history of items). DLRSs first map these categorical features into real-valued dense vectors via an~\textit{embedding-component}~\cite{pan2018field,qu2016product,zhou2018deep}, i.e., the embedding-lookup process, which leads to huge amounts of embedding parameters. For instance, the YouTube recommender system consists of 1 million of unique videoIDs, and assign each videoID with a specific 256-dimensional embedding vector; in other words, the videoID feature field alone occupies 256 million parameters~\cite{covington2016deep}. Then, the DLRSs nonlinearly transform the input embeddings from all feature fields and generate the outputs (predictions) via the \textit{MLP-component} (Multi-Layer Perceptron), which usually involves only several fully connected layers in practice. Therefore, compared to the MLP-component, the embedding-component dominates the number of parameters in practical recommender systems, which naturally plays a tremendously impactful role in the recommendations. 

The majority of existing recommender systems assign fixed and unified embedding dimension for all feature fields, such as the famous Wide\&Deep model~\cite{cheng2016wide}, which may lead to memory inefficiency. First, the embedding dimension often determines the capacity to encode information. Thus, allocating the same dimension to all feature fields may lose the information of high predictive features while wasting memory on non-predictive features. Therefore, we should assign large dimension to the high informative and predictive features, for instance, the ``location'' feature in location-based recommender systems~\cite{Bao2015Recommendations}. Second, different feature fields have different cardinality (i.e. the number of unique values). For example, the gender feature has only two (i.e. male and female), while the itemID feature usually involves millions of unique values. Intuitively, we should allocate larger dimensions to the feature fields with more unique feature values to encode their complex relationships with other features, and assign smaller dimensions to feature fields with smaller cardinality to avoid the overfitting problem due to the over-parameterization~\cite{joglekar2019neural,ginart2019mixed,zhao2020autoemb,kang2020learning}. According to the above reasons, it is highly desired to assign different embedding dimensions to different feature fields in a memory-efficient manner.

In this paper, we aim to enable different embedding dimensions for different feature fields for recommendations. We face tremendous challenges. First, the relationship among embedding dimensions, feature distributions and neural network architectures is highly intricate, which makes it hard to manually assign embedding dimensions to each feature field~\cite{ginart2019mixed}. Second, real-world recommender systems often involve hundreds and thousands of feature fields. It is difficult, if possible, to artificially select different dimensions for all feature fields, due to the expensive computation cost from the incredibly huge ($N^M$, with $N$ the number of candidate dimensions for each feature field to select, and $M$ the number of feature fields) search space. Our attempt to address these challenges results in an end-to-end differentiable AutoML based framework (\textbf{AutoDim}), which can efficiently allocate embedding dimensions to different feature fields in an automated and data-driven manner. Our experiments on benchmark datasets demonstrate the effectiveness of the proposed framework. We summarize our major contributions as: (i) we propose an end-to-end AutoML based framework AutoDim, which can automatically select various embedding dimensions to different feature fields; (ii) we develop two embedding lookup methods and two embedding transformation approaches, and compare the impact of their combinations on the embedding dimension allocation decision; and (iii) we demonstrate the effectiveness of the proposed framework on real-world benchmark datasets.

The rest of this paper is organized as follows. In Section 2, we introduce details about how to assign various embedding dimensions for different feature fields in an automated and data-driven fashion, and propose an AutoML based optimization algorithm. Section 3 carries out experiments based on real-world datasets and presents experimental results. Section 4 briefly reviews related work. Finally, section 5 concludes this work and discusses our future work.
\section{Framework}
\label{sec:framework}
In order to achieve the automated allocation of different embedding dimensions to different feature fields, we propose an AutoML based framework, which effectively addresses the challenges we discussed in Section~\ref{sec:introduction}. In this section, we will first introduce the overview of the whole framework; then we will propose an end-to-end differentiable model with two embedding-lookup methods and two embedding dimension search methods, which can compute the weights of different dimensions for feature fields in a soft and continuous fashion, and we will provide an AutoML based optimization algorithm; finally, we will derive a discrete embedding architecture upon the maximal weights, and retrain the whole DLRS framework.

\subsection{Overview}
\label{sec:Overview}
Our goal is to assign different feature fields various embedding dimensions in an automated and data-driven manner, so as to enhance the memory efficiency and the performance of the recommender system. We illustrate the overall framework in Figure~\ref{fig:Fig2_Framework}, which consists of two major stages: 

\subsubsection{\textbf{Dimensionality search stage}} 
It aims to find the optimal embedding dimension for each feature field. To be more specific, we first assign a set of candidate embeddings with different dimensions to a specific categorical feature via an \textit{embedding-lookup} step; then, we unify the dimensions of these candidate embeddings through a \textit{transformation} step, which is because of the fixed input dimension of the first MLP layer; next, we obtain the formal embedding for this categorical feature by computing the weighted sum of all its transformed candidate embeddings, and feed it into the MLP-component. The DLRS parameters including the embeddings and MLP layers are learned upon the training set, while the \textit{architectural weights} over the candidate embeddings are optimized upon the validation set, which prevents the framework selecting the embedding dimensions that overfit the training set~\cite{pham2018efficient,liu2018darts}. 

In practice, before the alternative training of DLRS parameters and architectural weights, we initially assign equivalent architectural weights on all candidate embeddings (e.g., $[0.5, 0.5]$ for the example in Figure~\ref{fig:Fig2_Framework}), fix these architectural weights and pre-train the DLRS including all candidate embeddings. The pre-training enables a fair competition between candidate embeddings when we start to update architectural weights.

\subsubsection{\textbf{Parameter re-training stage}} 
According to the \textit{architectural weights} learned in dimensionality search stage, we select the embedding dimension for each feature field, and re-train the parameters of DLRS parameters (i.e., MLPs and selected embeddings) on the training dataset in an end-to-end fashion. It is noteworthy that: (i) re-training stage is necessary, since in dimensionality search stage, the model performance is also influenced by the suboptimal embedding dimensions, which are not desired in practical recommender system; and (ii) new embeddings are still unified into the same dimension, since most existing deep recommender models (such as FM~\cite{rendle2010factorization}, DeepFM~\cite{guo2017deepfm}, NFM~\cite{he2017neural}) capture the interactions between two feature fields via a interaction operation (e.g., inner product) over their embedding vectors. These interaction operations constrain the embedding vectors to have same dimensions.

Note that \textit{numerical features} will be converted into categorical features through bucketing, and we omit this process in the following sections for simplicity. Next, we will introduce the details of each stage.

\begin{figure}[t]
    \centering
    \includegraphics[width=72mm]{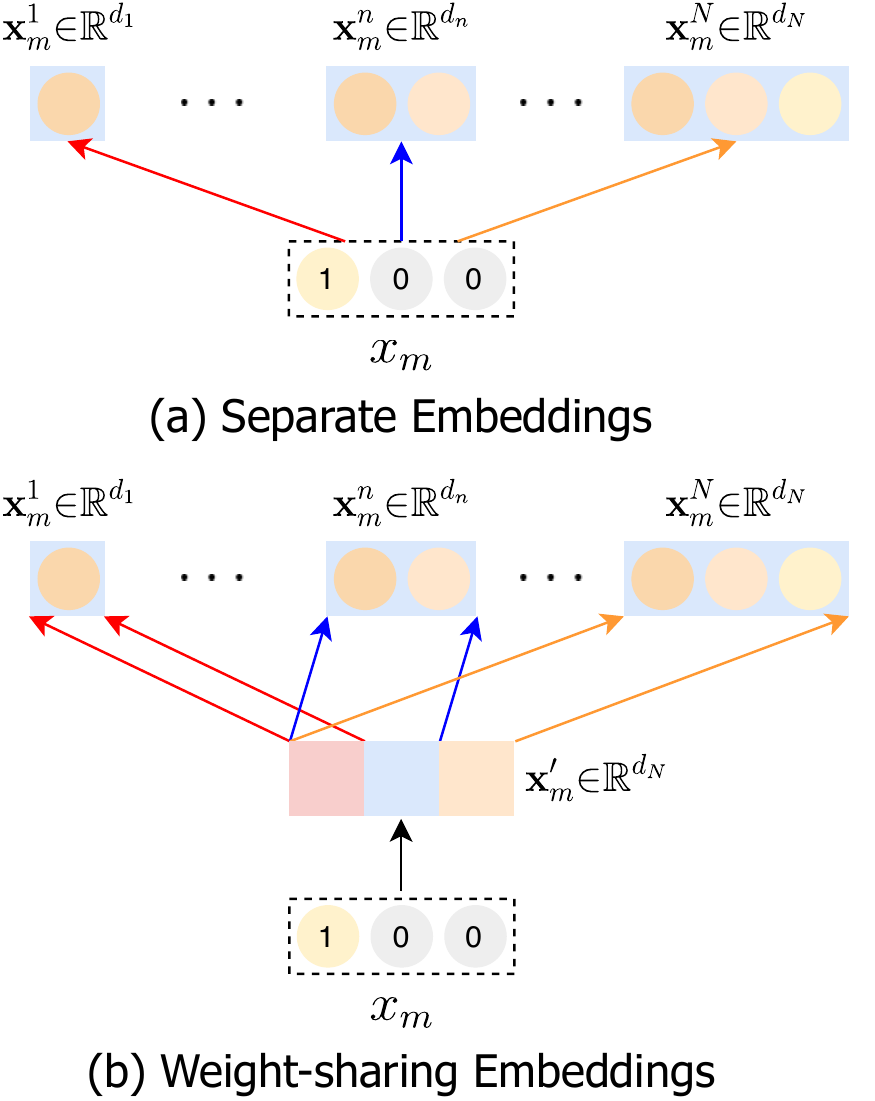}
    \caption{The embedding lookup methods.}
    \label{fig:Fig2_Lookup}
\end{figure}

\begin{figure*}[t]
	\centering
	\includegraphics[width=149mm]{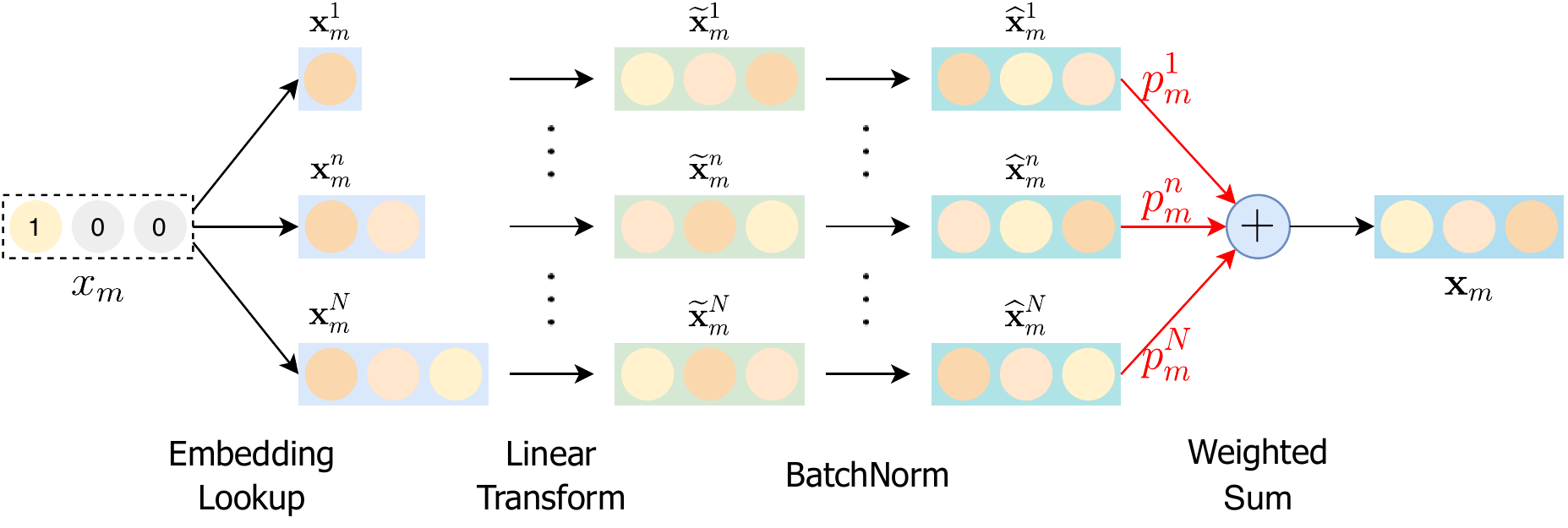}
	\caption{Method 1 - Linear Transformation.}
	\label{fig:Fig3_Search1}
\end{figure*}

\subsection{Dimensionality Search}
\label{sec:DimensionalitySearch}
As discussed in Section~\ref{sec:introduction}, different feature fields have different cardinalities and various contributions to the final prediction. Inspired by this phenomenon, it is highly desired to enable various embedding dimensions for different feature fields. However, due to a large amount of feature fields and the complex relationship between embedding dimensions with feature distributions and neural network architectures, it is difficult to manually select embedding dimensions via conventional dimension reduction methods. An intuitive solution to tackle this challenge is to assign several embedding spaces with various dimensions to each feature field, and then the DLRS automatically selects the optimal embedding dimension for each feature field.

\subsubsection{\textbf{Embedding Lookup Tricks}} 
\label{sec:Lookup}
Suppose for each user-item interaction instance, we have $M$ input features $(x_1,\cdots,x_M)$, and each feature $x_m$ belongs to a specific feature field, such as gender and age, etc. For the $m^{th}$ feature field, we assign $N$ candidate embedding spaces $\{\mathbf{X}_m^1,\cdots,\mathbf{X}_m^N\}$. The dimension of an embedding in each space is $d_1,\cdots,d_N$, where $d_1\textless\cdots \textless d_N$; and the cardinality of these embedding spaces are the number of unique feature values in this feature field. Correspondingly, we define $\{\mathbf{x}_m^1,\cdots,\mathbf{x}_m^N\}$ as the set of candidate embeddings for a given feature $x_m$ from all embedding spaces, as shown in Figure \ref{fig:Fig2_Lookup} (a). Note that we assign the same candidate dimension to all feature fields for simplicity, but it is straightforward to introduce different candidate sets. Therefore, the total space assigned to the feature $x_m$ is $\sum_{n=1}^{N} d_n$. However, in real-world recommender systems with thousands of feature fields, two challenges lie in this design include (i) this design needs huge space to store all candidate embeddings, and (ii) the training efficiency is reduced since a large number of parameters need to be learned. 

To address these challenges, we propose an alternative solution for large-scale recommendations, named \textit{weight-sharing embedding} architecture. As illustrated in Figure \ref{fig:Fig2_Lookup} (b), we only allocate a $d_N$-dimensional embedding to a given feature $x_m$, referred as to $\mathbf{x}_m'$, then the $n^{th}$ candidate embedding $\mathbf{x}_m^n$ corresponds to the first $d_n$ digits of $\mathbf{x}_m'$. The advantages associated with weight-sharing embedding method are two-fold, i.e., (i) it is able to reduce the storage space and increase the training efficiency, as well as (ii) since the relatively front digits of $\mathbf{x}_m'$ have more chances to be retrieved and then be trained (e.g. the ``red part'' of $\mathbf{x}_m'$ is leveraged by all candidates in Figure \ref{fig:Fig2_Lookup} (b)), we intuitively wish they can capture more essential information of the feature $x_m$.

\subsubsection{\textbf{Unifying Various Dimensions}}
\label{sec:Unifying}
Since the input dimension of the first MLP layer in existing DLRSs is often fixed, it is difficult for them to handle various candidate dimensions. Thus we need to unify the embeddings $\{\mathbf{x}_m^1,\cdots,\mathbf{x}_m^N\}$ into same dimension, and we develop two following methods:

\paragraph{\textbf{Method 1: Linear Transformation}} 
Figure~\ref{fig:Fig3_Search1} (a) illustrates the linear transformation method to handle the various embedding dimensions (the difference of two embedding lookup methods is omitted here). We introduce $N$ fully-connected layers, which transform embedding vectors $\{\mathbf{x}_m^1,\cdots,\mathbf{x}_m^N\}$ into the same dimension $d_N$: 
\begin{equation}
    \label{equ:Linear}
    \begin{array}{l}
        {\mathbf{\widetilde{x}}_m^n\leftarrow\mathbf{W}_{n}^{\top} \mathbf{x}_m^n+\mathbf{b}_{n}} \quad \forall n \in[1,N]
    \end{array}
\end{equation}
\noindent where $\mathbf{W}_{n}\in \mathbb{R}^{d_n \times d_N}$ is weight matrice and $\mathbf{b}_{n}\in \mathbb{R}^{d_N}$ is bias vector. For each field, all candidate embeddings with the same dimension share the same weight matrice and bias vector, which can reduce the amount of model parameters. With the linear transformations, we map the original embedding vectors $\{\mathbf{x}_m^1,\cdots,\mathbf{x}_m^N\}$ into the same dimensional space, i.e., $\{\mathbf{\widetilde{x}}_m^1,\cdots,\mathbf{\widetilde{x}}_m^N\}\in \mathbb{R}^{d_N}$. In practice, we can observe that the magnitude of the transformed embeddings $\{\mathbf{\widetilde{x}}_m^1,\cdots,\mathbf{\widetilde{x}}_m^N\}$ varies significantly, which makes them become incomparable. To tackle this challenge, we conduct BatchNorm~\cite{ioffe2015batch} on the transformed embeddings $\{\mathbf{\widetilde{x}}_m^1,\cdots,\mathbf{\widetilde{x}}_m^N\}$ as:
\begin{equation}
    \label{equ:BatchNorm}
    \begin{array}{l}
        \mathbf{\widehat{x}}_m^n \leftarrow \frac{\mathbf{\widetilde{x}}_m^n-\mu_{\mathcal{B}}^n}{\sqrt{(\sigma_{\mathcal{B}}^{n})^2+\epsilon}} \quad \forall n \in[1,N]
    \end{array}
\end{equation}
\noindent where $\mu_{\mathcal{B}}^n$ is the mini-batch mean and $(\sigma_{\mathcal{B}}^{n})^2$ is the mini-batch variance for $\forall n \in [1,N]$. $\epsilon$ is a small constant
added to the mini-batch variance for numerical stability when $(\sigma_{\mathcal{B}}^{n})^2$ is very small. After BatchNorm, the linearly transformed embeddings $\{\mathbf{\widetilde{x}}_m^1,\cdots,\mathbf{\widetilde{x}}_m^N\}$ become to magnitude-comparable embedding vectors $\{\mathbf{\widehat{x}}_m^1,\cdots,\mathbf{\widehat{x}}_m^N\}$ with the same dimension $d_N$.

\begin{figure*}[t]
    \centering
    \includegraphics[width=149mm]{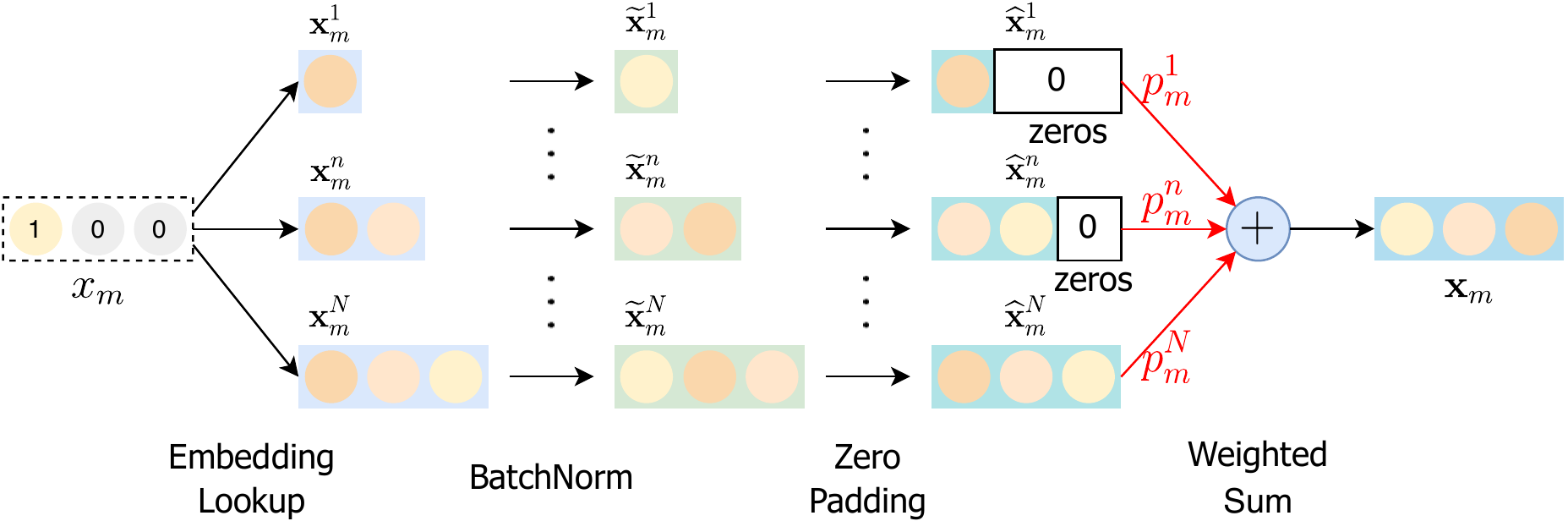}
    \caption{Method 2 - Zero Padding Transformation.}
    \label{fig:Fig3_Search2}
\end{figure*}

\paragraph{\textbf{Method 2: Zero Padding}} 
Inspired by zero-padding techniques from the computer version community, which pads the input volume with zeros around the border, we address the problem of various embedding dimensions by padding shorter embedding vectors to the same length as the longest embedding dimension $d_N$ with zeros, which is illustrated in Figure~\ref{fig:Fig3_Search2}. For the embedding vectors $\{\mathbf{x}_i^1,\cdots,\mathbf{x}_i^N\}$ with different dimensions, we first execute BatchNorm process, which forces the original  embeddings $\{\mathbf{x}_i^1,\cdots,\mathbf{x}_i^N\}$ into becoming magnitude-comparable embeddings:
\begin{equation}
    \label{equ:BatchNorm1}
    \begin{array}{l}
        \mathbf{\widetilde{x}}_m^n \leftarrow \frac{\mathbf{x}_m^n-\mu_{\mathcal{B}}^n}{\sqrt{(\sigma_{\mathcal{B}}^{n})^2+\epsilon}} \quad \forall n \in[1,N]
    \end{array}
\end{equation}
\noindent where $\mu_{\mathcal{B}}^n$, $(\sigma_{\mathcal{B}}^{n})^2$ are the mini-batch mean and variance. $\epsilon$ is the constant for numerical stability. The transformed $\{\mathbf{\widetilde{x}}_m^1,\cdots,\mathbf{\widetilde{x}}_m^N\}$ are magnitude-comparable embeddings. Then we pad the $\{\mathbf{\widetilde{x}}_m^1,\cdots,\mathbf{\widetilde{x}}_m^{N-1}\}$ to the same length $d_N$ by zeros: 
\begin{equation}
    \label{equ:Padding}
    \begin{array}{l}
        \,\,\,\,\,{\mathbf{\widehat{x}}_m^n\leftarrow padding(\,\mathbf{\widetilde{x}}_m^n\,,\,d_N-d_n)} \quad \forall n \in[1,N]
    \end{array}
\end{equation}
\noindent where the second term of each padding formula is the number of zeros to be padded with the embedding vector of the first term. Then the embeddings $\{\mathbf{\widehat{x}}_m^1,\cdots,\mathbf{\widehat{x}}_m^N\}$ share the same dimension $d_N$. Compared with the linear transformation (method 1), the zero-padding method reduces lots of linear-transformation computations and corresponding parameters. The possible drawback is that the final embeddings $\{\mathbf{\widehat{x}}_m^1,\cdots,\mathbf{\widehat{x}}_m^N\}$ becomes spatially unbalanced since the tail parts of some final embeddings are zeros. Next, we will introduce embedding dimension selection process.

\subsubsection{\textbf{Dimension Selection}}
In this paper, we aim to select the optimal embedding dimension for each feature field in an automated and data-driven manner. This is a hard (categorical) selection on the candidate embedding spaces, which will make the whole framework not end-to-end differentiable. To tackle this challenge, in this work, we approximate the hard selection over different dimensions via introducing the Gumbel-softmax operation~\cite{jang2016categorical}, which simulates the non-differentiable sampling from a categorical distribution by a differentiable sampling from the Gumbel-softmax distribution.

To be specific, suppose weights $\{\alpha_m^1, \cdots, \alpha_m^N\}$ are the class probabilities over different dimensions. Then a hard selection $z$ can be drawn via the the gumbel-max trick~\cite{gumbel1948statistical} as: 
\begin{equation}
    \begin{aligned}
        z=& \;\text{one\_hot} \left(\arg \max_{n\in[1,N]}\left[\log\alpha_m^n+g_{n}\right]\right)\\
        &where\;g_{n}=-\log \left(-\log \left(u_{n}\right)\right)\\
        &\quad\quad\;\;\, u_{n} \sim Uniform (0,1)
    \end{aligned}
\end{equation}
\noindent The \textit{gumbel noises} $g_{i},\cdots, g_{N}$ are i.i.d samples, which perturb $\log\alpha_m^n$ terms and make the $\arg \max$ operation that is equivalent to drawing a sample by $\alpha_m^1, \cdots, \alpha_m^N$ weights. However, this trick is non-differentiable due to the $\arg \max$ operation. To deal with this problem, we use the softmax function as a continuous, differentiable approximation to $\arg \max$ operation, i.e., straight-through gumbel-softmax~\cite{jang2016categorical}: 
\begin{equation}
\label{equ:p_m^n}
    p_m^n=\frac{\exp \left(\frac{\log \left(\alpha_m^n\right)+g_{n}}{\tau}\right)}{\sum_{i=1}^{N} \exp \left(\frac{\log \left(\alpha_m^i\right)+g_{i}}{\tau}\right)}
\end{equation}
\noindent where $\tau$ is the temperature parameter, which controls the smoothness of the output of gumbel-softmax operation. When $\tau$ approaches zero, the output of the gumbel-softmax becomes closer to a one-hot vector. Then $p_m^n$ is the probability of selecting the $n^{th}$ candidate embedding dimension for the feature $x_m$, and its embedding $\mathbf{x}_m$ can be formulated as the weighted sum of $\{\mathbf{\widehat{x}}_m^1,\cdots,\mathbf{\widehat{x}}_m^N\}$: 
\begin{equation}
	\begin{array}{l}
		\label{equ:WeightedSum}
       \mathbf{x}_m = \sum_{n=1}^{N} p_m^n\cdot\mathbf{\widehat{x}}_m^n \quad \forall m \in [1,M]
	\end{array}
\end{equation}
We illustrate the weighted sum operations in Figure~\ref{fig:Fig3_Search1} and \ref{fig:Fig3_Search2}.
With gumbel-softmax operation, the dimensionality search process is end-to-end differentiable. The discrete embedding dimension selection conducted based on the weights $\{\alpha_m^n\}$ will be detailed in the following subsections. 

Then, we concatenate the embeddings $\mathbf{h}_{0}=\left[\mathbf{x}_1,\cdots,\mathbf{x}_M\right]$ and feed $\mathbf{h}_{0}$ input into $L$ multilayer perceptron layers:
\begin{equation}
    \begin{array}{l}
    {\mathbf{h}_{l}=\sigma\left(\mathbf{W}_{l}^{\top} \mathbf{h}_{l-1}+\mathbf{b}_{l}\right)}  \quad \forall l \in [1,L]
    \end{array}
\end{equation}
\noindent where $\mathbf{W}_{l}$ and $\mathbf{b}_{l}$ are the weight matrix and the bias vector for the $l^{th}$ MLP layer. $\sigma(\cdot)$ is the activation function such as \textit{ReLU} and \textit{Tanh}. Finally, the output layer that is subsequent to the last MLP layer, produces the prediction of the current user-item interaction instance as:
\begin{equation}
    {\hat{y}=\sigma\left(\mathbf{W}_{o}^{\top} \mathbf{h}_{L}+\mathbf{b}_{o}\right)}
\end{equation}
\noindent where $\mathbf{W}_{o}$ and $\mathbf{b}_{o}$ are the  weight matrix and bias vector for the output layer. Activation function $\sigma(\cdot)$ is selected based on different recommendation tasks, such as Sigmoid function for regression~\cite{cheng2016wide}, and Softmax for multi-class classification~\cite{tan2016improved}. Correspondingly, the objective function $\mathcal{L}(\hat{y}, y)$ between prediction $\hat{y}$ and ground truth label $y$ also varies based on different recommendation tasks. In this work, we leverage negative log-likelihood function: 
\begin{equation}
    \label{equ:loss}
    \mathcal{L}(\hat{y}, y)=-y \log \hat{y}-(1-y) \log (1-\hat{y})
\end{equation}
\noindent where $y$ is the ground truth (1 for like or click, 0 for dislike or non-click). By minimizing the objective function $\mathcal{L}(\hat{y}, y)$, the dimensionality search framework updates the parameters of all embeddings, hidden layers, and weights $\{\alpha_m^n\}$ through back-propagation. The high-level idea of the dimensionality search is illustrated in Figure~\ref{fig:Fig2_Framework} (a), where we omit some details of embedding-lookup, transformations and gumbel-softmax for the sake of simplicity.

\subsection{Optimization}
\label{sec:Optimization}
In this subsection, we will detail the optimization method of the proposed AutoDim framework. In AutoDim, we formulate the selection over different embedding dimensions as an architectural optimization problem and make it end-to-end differentiable by leveraging the Gumbel-softmax technique. The parameters to be optimized in AutoDim are two-fold, i.e.,  (i) $\mathbf{W}$: the parameters of the DLRS, including the embedding-component and the MLP-component; (ii) $\boldsymbol{\alpha}$: the weights $\{\alpha_m^n\}$ on different embedding spaces ($\{p_m^n\}$ are calculated based on $\{\alpha_m^n\}$ as in Equation (\ref{equ:p_m^n})). DLRS parameters $\mathbf{W}$ and architectural weights $\boldsymbol{\alpha}$ can not be optimized simultaneously on training dataset as conventional supervised attention mechanism since the optimization of them are highly dependent on each other. In other words, simultaneously optimization on training dataset may result in model overfitting on the examples from training dataset. 

Inspired by the differentiable architecture search (DARTS) techniques~\cite{liu2018darts}, $\mathbf{W}$ and $\boldsymbol{\alpha}$ are alternately optimized through gradient descent. Specifically, we alternately update $\mathbf{W}$ by optimizing the loss $\mathcal{L}_{train}$ on the training data and update $\boldsymbol{\alpha}$ by optimizing the loss $\mathcal{L}_{val}$ on the validation data:
\begin{equation}
\begin{aligned}
\label{equ:bilevel}
\min_{\boldsymbol{\alpha}} \; &\mathcal{L}_{val} \big(\mathbf{W}^*(\boldsymbol{\alpha}),\boldsymbol{\alpha}\big)\\
s.t. \; & \mathbf{W}^*(\boldsymbol{\alpha}) = \arg\min_\mathbf{W} \mathcal{L}_{train} (\mathbf{W}, \boldsymbol{\alpha}^*)
\end{aligned}
\end{equation}
\noindent this optimization forms a bilevel optimization problem~\cite{pham2018efficient}, where architectural weights $\boldsymbol{\alpha}$ and DLRS parameters $\mathbf{W}$ are identified as the upper-level variable and lower-level variable. Since the inner optimization of $\mathbf{W}$ is computationally expensive, directly optimizing $\boldsymbol{\alpha}$ via Eq.(\ref{equ:bilevel}) is intractable. To address this challenge, we take advantage of the approximation scheme of DARTS:
\begin{algorithm}[t]
    \caption{\label{alg:DARTS} DARTS based Optimization for AutoDim.}
    \raggedright
    {\bf Input}: the features $(x_1,\cdots,x_M)$ of user-item interactions and the corresponding ground-truth labels $y$\\
    {\bf Output}: the well-learned DLRS parameters $\mathbf{W}^*$; the well-learned weights on various embedding spaces $\boldsymbol{\alpha}^*$\\
    \begin{algorithmic} [1]
        \WHILE{not converged}
        \STATE Sample a mini-batch of user-item interactions from validation data
        \STATE Update $\boldsymbol{\alpha}$ by descending $\nabla_{\boldsymbol{\alpha}} \;\mathcal{L}_{val} \big(\mathbf{W}^*(\boldsymbol{\alpha}),\boldsymbol{\alpha}\big)$ with the approximation in Eq.(\ref{equ:approximation})
        \STATE Collect a mini-batch of training data
        \STATE Generate predictions $\hat{y}$ via DLRS with current $\mathbf{W}$ and architectural weights $\boldsymbol{\alpha}$
        \STATE Update $\mathbf{W}$ by descending $\nabla_{\mathbf{W}}\mathcal{L}_{train} (\mathbf{W}, \boldsymbol{\alpha})$
        \ENDWHILE
    \end{algorithmic}
\end{algorithm}
\begin{equation}
\begin{aligned}
\label{equ:approximation}
& \arg\min_\mathbf{W} \mathcal{L}_{train} (\mathbf{W}, \boldsymbol{\alpha}^*) \approx \mathbf{W} - \xi \nabla_{\mathbf{W}}\mathcal{L}_{train} (\mathbf{W}, \boldsymbol{\alpha})
\end{aligned}
\end{equation}
\noindent where $\xi$ is the learning rate. In the approximation scheme, when updating $\boldsymbol{\alpha}$ via Eq.(\ref{equ:approximation}), we estimate $\mathbf{W}^*(\boldsymbol{\alpha})$ by descending the gradient $\nabla_{\mathbf{W}}\mathcal{L}_{train} (\mathbf{W}, \boldsymbol{\alpha})$ for only one step, rather than to optimize $\mathbf{W}(\boldsymbol{\alpha})$ thoroughly to obtain $\mathbf{W}^*(\boldsymbol{\alpha}) = \arg\min_\mathbf{W}$ $ \mathcal{L}_{train} (\mathbf{W}, \boldsymbol{\alpha}^*)$. In practice, it usually  leverages the first-order approximation by setting $\xi=0$, which can further enhance the computation efficiency. 

The DARTS based optimization algorithm for AutoDim is detailed in Algorithm \ref{alg:DARTS}. Specifically, in each iteration, we first sample a batch of user-item interaction data from the validation set (line 2); next, we update the architectural weights  $\boldsymbol{\alpha}$ upon it (line 3); afterward, the DLRS make the predictions $\hat{y}$ on the batch of training data with current DLRS parameters $\mathbf{W}$ and architectural weights $\boldsymbol{\alpha}$ (line 5); eventually, we update the DLRS parameters $\mathbf{W}$ by descending $\nabla_{\mathbf{W}}\mathcal{L}_{train} (\mathbf{W}, \boldsymbol{\alpha})$ (line 6).

\subsubsection{A pre-train trick.} In practice, in order to enable a fair competition between the candidate embeddings, for each feature field, we first allocate the equivalent architectural weights initially on all its candidate embeddings, e.g., $[0.5, 0.5]$ if there are two candidate embedding dimensions. Then, we fix these initialized architectural weights $\boldsymbol{\alpha}$ and pre-train the DLRS parameters $\mathbf{W}$ including all candidate embeddings. This process ensures a fair competition between candidate embeddings when we begin to update $\boldsymbol{\alpha}$.

 \subsection{Parameter Re-Training}
Since the suboptimal embedding dimensions in dimensionality search stage also influence the model training, a re-training stage is desired to training the model with only optimal dimensions, which can eliminate these suboptimal influences. In this subsection, we will introduce how to select optimal embedding dimension for each feature field and the details of re-training the recommender system with the selected embedding dimensions. 

\subsubsection{\textbf{Deriving Discrete Dimensions}}
\label{sec:Deriving}
During re-training, the gumbel-softmax operation is no longer used, which means that the optimal embedding space (dimension) are selected for each feature field as the one corresponding to the largest weight, based on the well-learned $\boldsymbol{\alpha}$. It is formally defined as: 
\begin{equation}
    \label{equ:Hardselection1}
    \begin{array}{l}
        \;\mathbf{X}_m = \mathbf{X}_m^{k} ,\;\;\; where \;\; k = \arg\max_{n \in [1,N]} \alpha_m^n \quad \forall m \in [1,M]
    \end{array}
\end{equation}
Figure~\ref{fig:Fig2_Framework} (a) illustrates the architecture of AutoDim framework with a toy example about the optimal dimension selections based on two candidate dimensions, where the largest weights corresponding to the $1^{st}$, $m^{th}$ and $M^{th}$ feature fields are  $0.7$, $0.8$ and $0.6$, then the embedding space $\mathbf{X}_1^{1}$, $\mathbf{X}_m^{2}$ and $\mathbf{X}_M^{1}$ are selected for these feature fields. The dimension of an embedding vector in these embedding spaces is $d_1$, $d_2$ and $d_1$, respectively.

\begin{algorithm}[t]
	\caption{\label{alg:re-training} The Optimization of DLRS Re-training Process.}
	\raggedright
	{\bf Input}: the features $(x_1,\cdots,x_M)$ of user-item interactions and the corresponding ground-truth labels $y$\\
	{\bf Output}: the well-learned DLRS parameters $\mathbf{W}^*$\\
	\begin{algorithmic} [1]
		\WHILE{not converged}
		\STATE Sample a mini-batch of training data
		\STATE Generate predictions $\hat{y}$ via DLRS with current  $\mathbf{W}$
		\STATE Update $\mathbf{W}$ by descending $\nabla_{\mathbf{W}}\mathcal{L}_{train} (\mathbf{W})$
		\ENDWHILE
	\end{algorithmic}
\end{algorithm}

\subsubsection{\textbf{Model Re-training}} 
As shown in Figure~\ref{fig:Fig2_Framework} (b), given the selected embedding spaces, we can obtain unique embedding vectors $(\mathbf{x}_1,\cdots,\mathbf{x}_M)$ for features $(x_1,\cdots,x_M)$. Then we concatenate these embeddings and feeds them into hidden layers. Next, the prediction $\hat{y}$ is generated by the output layer. Finally, all the parameters of the DLRS, including embeddings and MLPs, will be updated via minimizing the supervised loss function $\mathcal{L}(\hat{y}, y)$  through back-propagation. The model re-training algorithm is detailed in Algorithm~\ref{alg:re-training}. The re-training process is based on the same training data as Algorithm~\ref{alg:DARTS}.

Note that the majority of existing deep recommender algorithms (such as FM~\cite{rendle2010factorization}, DeepFM~\cite{guo2017deepfm}, FFM~\cite{pan2018field}, AFM~\cite{xiao2017attentional}, xDeepFM~\cite{lian2018xdeepfm}) capture the interactions between feature fields via interaction operations, such as inner product and Hadamard product. These interaction operations require the embedding vectors from all fields to have the same dimensions. Therefore, the embeddings selected in Section~\ref{sec:Deriving} are still mapped into the same dimension as in Section \ref{sec:Unifying}. In the re-training stage, the BatchNorm operation is no longer in use, since there are no comparisons between candidate embeddings in each field. Unifying embeddings into the same dimension does not increase model parameters and computations too much: (i) linear transformation: all embeddings from one feature field share the same weight matrice and bias vector, and (ii) zero-padding: no extra trainable parameters are introduced.
\section{Experiments}
\label{sec:experiments}
In this section, we first introduce experimental settings. Then we conduct extensive experiments to evaluate the effectiveness of the proposed AutoDim framework. We mainly seek answers to the following research questions:
\textbf{RQ1}: How does AutoDim perform compared with other embedding dimension search methods?
\textbf{RQ2}: How do the important components, i.e., 2 embedding lookup methods and 2 transformation methods, influence the performance?
\textbf{RQ3}: How efficient is AutoDim as compared with other methods?
\textbf{RQ4}: What is the impact of important parameters on the results?
\textbf{RQ5}: What is the transferability and stability of AutoDim?
\textbf{RQ6}: Can AutoDim assign large embedding dimensions to really important feature fields?

\begin{table}[]
	\caption{Statistics of the datasets.}
	\label{table:statistics}
	\begin{tabular}{@{}|c|c|c|@{}}
		\toprule[1pt]
		Data & Criteo & Avazu \\ \midrule
		\# Interactions & 45,840,617 & 40,428,968 \\
		\# Feature Fields & 39 & 22 \\
		\# Sparse Features & 1,086,810 & 2,018,012 \\ \bottomrule[1pt]
	\end{tabular}
\end{table}

\begin{table*}[]
	\caption{Performance comparison of different embedding search methods}
	\label{table:result1}
	\begin{tabular}{@{}|c|c|c|ccccccccc|@{}}
		\toprule[1pt]
		\multirow{2}{*}{Dataset} & \multirow{2}{*}{Model} & \multirow{2}{*}{Metrics} & \multicolumn{9}{c|}{Search Methods} \\ \cmidrule(l){4-12} 
		&  &  & FDE & MDE & DPQ & NIS & MGQE & AutoEmb & RaS & AutoDim-s & AutoDim \\ \midrule
		\multirow{3}{*}{Criteo} & \multirow{3}{*}{FM} & AUC & 0.8020 & 0.8027 & 0.8035 & 0.8042 & 0.8046 & 0.8049 & 0.8056 & 0.8063 & \textbf{0.8078*} \\
		&  & Logloss & 0.4487 & 0.4481 & 0.4472 & 0.4467 & 0.4462 & 0.4460 & 0.4457 & 0.4452 & \textbf{0.4438*} \\
		&  & Params (M) & 34.778 & 15.520 & 20.078 & 13.636 & 12.564 & 13.399 & 16.236 & 31.039 & \textbf{11.632*} \\ \midrule
		\multirow{3}{*}{Criteo} & \multirow{3}{*}{W\&D} & AUC & 0.8045 & 0.8051 & 0.8058 & 0.8067 & 0.8070 & 0.8072 & 0.8076 & 0.8081 & \textbf{0.8098*} \\
		&  & Logloss & 0.4468 & 0.4464 & 0.4457 & 0.4452 & 0.4446 & 0.4445 & 0.4443 & 0.4439 & \textbf{0.4419*} \\
		&  & Params (M) & 34.778 & 18.562 & 22.628 & 14.728 & 15.741 & 15.987 & 18.233 & 30.330 & \textbf{12.455*} \\ \midrule
		\multirow{3}{*}{Criteo} & \multirow{3}{*}{DeepFM} & AUC & 0.8056 & 0.8060 & 0.8067 & 0.8076 & 0.8080 & 0.8082 & 0.8085 & 0.8089 & \textbf{0.8101*} \\
		&  & Logloss & 0.4457 & 0.4456 & 0.4449 & 0.4442 & 0.4439 & 0.4438 & 0.4436 & 0.4432 & \textbf{0.4416*} \\
		&  & Params (M) & 34.778 & 17.272 & 25.737 & 12.955 & 13.059 & 13.437 & 17.816 & 31.770 & \textbf{11.457*} \\ \midrule
		\multirow{3}{*}{Avazu} & \multirow{3}{*}{FM} & AUC & 0.7799 & 0.7802 & 0.7809 & 0.7818 & 0.7823 & 0.7825 & 0.7827 & 0.7831 & \textbf{0.7842*} \\
		&  & Logloss & 0.3805 & 0.3803 & 0.3799 & 0.3792 & 0.3789 & 0.3788 & 0.3787 & 0.3785 & \textbf{0.3776*} \\
		&  & Params (M) & 64.576 & 22.696 & 28.187 & 22.679 & 22.769 & 21.026 & 27.272 & 55.038 & \textbf{17.595*} \\ \midrule
		\multirow{3}{*}{Avazu} & \multirow{3}{*}{W\&D} & AUC & 0.7827 & 0.7829 & 0.7836 & 0.7842 & 0.7849 & 0.7851 & 0.7853 & 0.7856 & \textbf{0.7872*} \\
		&  & Logloss & 0.3788 & 0.3785 & 0.3777 & 0.3772 & 0.3768 & 0.3767 & 0.3767 & 0.3766 & \textbf{0.3756*} \\
		&  & Params (M) & 64.576 & 27.976 & 35.558 & 21.413 & 19.457 & 17.292 & 35.126 & 56.401 & \textbf{14.130*} \\ \midrule
		\multirow{3}{*}{Avazu} & \multirow{3}{*}{DeepFM} & AUC & 0.7842 & 0.7845 & 0.7852 & 0.7858 & 0.7863 & 0.7866 & 0.7867 & 0.7870 & \textbf{0.7881*} \\
		&  & Logloss & 0.3742 & 0.3739 & 0.3737 & 0.3736 & 0.3734 & 0.3733 & 0.3732 & 0.3730 & \textbf{0.3721*} \\
		&  & Params (M) & 64.576 & 32.972 & 36.128 & 22.550 & 17.575 & 21.605 & 29.235 & 58.325 & \textbf{13.976*} \\ \bottomrule[1pt]
	\end{tabular}
\\``\textbf{{\Large *}}" indicates the statistically significant improvements (i.e., two-sided t-test with $p<0.05$) over the best baseline. (M=Million)
\end{table*}

\subsection{Datasets}
\label{sec:Datasets}
We evaluate our model on two benchmark datasets:
(i) \textbf{Criteo}\footnote{https://www.kaggle.com/c/criteo-display-ad-challenge/}: This is a benchmark industry dataset to evaluate ad click-through rate prediction models. It consists of 45 million users' click records on displayed ads over one month. For each data example, it contains 13 numerical feature fields and 26 categorical feature fields. We normalize numerical features by transforming a value $v \rightarrow\left\lfloor\log (v)^{2}\right\rfloor$ if $v>2$ as proposed by the Criteo Competition winner \footnote{https://www.csie.ntu.edu.tw/~r01922136/kaggle-2014-criteo.pdf}, and then convert it into categorical features through bucketing. All $M=39$ feature fields are anonymous. 
(ii) \textbf{Avazu}\footnote{https://www.kaggle.com/c/avazu-ctr-prediction/}: Avazu dataset was provided for the CTR prediction challenge on Kaggle, which contains 11 days' user clicking behaviors that whether a displayed mobile ad impression is clicked or not. There are $M=22$ categorical feature fields including user/ad features and device attributes. Parts of the fields are anonymous.
Some key statistics of the datasets are shown in Table \ref{table:statistics}. For each dataset, we use 90\% user-item interactions as the training/validation set (8:1), and the rest 10\% as the test set.

\subsection{Implement Details}
\label{sec:architecture}
Next, we detail the AutoDim architectures. For the DLRS, (i) embedding component: existing work usually set the embedding dimension as 10 or 16, while recent research found that a larger embedding size leads to better performance~\cite{zhu2020fuxictr}, so we set the maximal embedding dimension as 32 within our GPU memory constraints. For each feature field, we select from $N=5$ candidate embedding dimensions $\{2,8,16,24,32\}$. 
(ii) MLP component: we have two hidden layers with the size $|h_0|\times 128$ and $128\times128$, where $|h_0|$ is the input size of first hidden layer, $|h_0|=32\times M$  with $M$ the number of feature fields for different datasets, and we use batch normalization, dropout ($rate=0.2$) and ReLU activation for both hidden layers. The output layer is $128\times1$ with Sigmoid activation. 

For architectural weights $\boldsymbol{\alpha}$: $\alpha_m^1, \cdots, \alpha_m^N$ of the $m^{th}$ feature field are produced by a Softmax activation upon a trainable vector of length $N$. We use an annealing temperature $\tau=\max (0.01, 1 - 0.00005 \cdot t)$ for Gumbel-softmax, where $t$ is the training step. 

The learning rate for updating DLRS and weights are $0.001$ and $0.001$, and the batch-size is set as 2000. For the parameters of the proposed AutoDim framework, we select them via cross-validation. Correspondingly, we also do parameter-tuning for baselines for a fair comparison. We will discuss more details about parameter selection for the proposed framework in the following subsections.

Our implementation is based on a public Pytorch for recommendation library\footnote{https://github.com/rixwew/pytorch-fm}, which involves 16 state-of-the-art recommendation algorithms. Our model is implemented as 3 separate classes/functions, so it is easily to be apply our AutoDim model to these recommendation algorithms. Due to the limited space, we show the performances of applying AutoDim on FM~\cite{rendle2010factorization}, W\&D~\cite{cheng2016wide} and DeepFM~\cite{guo2017deepfm}.

\subsection{Evaluation Metrics}
\label{sec:metrics}
The performance is evaluated by AUC, Logloss and Params, where a higher AUC or a lower Logloss indicates a better recommendation performance. A lower Params means a fewer embedding parameters. Area Under the ROC Curve (AUC) measures the probability that a positive instance will be ranked higher than a randomly chosen negative one; 
we introduce Logoss since all methods aim to optimize the logloss in Equation (\ref{equ:loss}), thus it is natural to utilize Logloss as a straightforward metric. It is noteworthy that a slightly higher AUC or lower Logloss at \textbf{0.001-level} is regarded as significant for the CTR prediction task~\cite{cheng2016wide,guo2017deepfm}. For an embedding dimension search model, the Params metric is the optimal number of embedding parameters selected by this model for the recommender system. We omit the number of MLP parameters, which only occupy a small part of the total model parameters, e.g., $\sim0.5\%$ in W\&D and DeepFM on Criteo dataset. FM model has  no MLP component.

\subsection{Overall Performance (RQ1)}
\label{sec:RQ1}
We compare the proposed framework with following embedding dimension search methods:
(i) \textbf{FDE} (Full Dimension Embedding): In this baseline, we assign the same embedding dimensions to all feature fields. For each feature field, the embedding dimension is set as the maximal size from the candidate set, i.e., 32.
(ii) \textbf{MDE} (Mixed Dimension Embedding)~\cite{ginart2019mixed}: This is a heuristic method that assigns highly-frequent feature values with larger embedding dimensions, vice versa. We enumerate its 16 groups of suggested hyperparameters settings and report the best one. 
(iii) \textbf{DPQ} (Differentiable Product Quantization)~\cite{chen2019differentiable}:  This baseline introduces differentiable quantization techniques from \textit{network compression} community to compact embeddings.
(iv) \textbf{NIS} (Neural Input Search)~\cite{joglekar2020neural}: This baseline applies reinforcement learning to learn to allocate larger embedding sizes to active feature values, and smaller sizes to inactive ones.
(v) \textbf{MGQE} (Multi-granular quantized embeddings)~\cite{kang2020learning}:  This baseline is based on DPQ, and further cuts down the embeddings space by using fewer centroids for non-frequent feature values.
(vi) \textbf{AutoEmb} (Automated Embedding Dimensionality Search)~\cite{zhao2020autoemb}:  This baseline is based on DARTS~\cite{liu2018darts}, and assigns embedding dimensions according to the frequencies of feature values. 
(vii) \textbf{RaS} (Random Search): Random search is strong baseline in neural network search~\cite{liu2018darts}. We apply the same candidate embedding dimensions, randomly allocate dimensions to feature fields in each experiment time, and report the best performance.
(viii) \textbf{AutoDim-s}: This baseline shares the same architecture with AutoDim, while we update the DLRS parameters and architectural weights simultaneously on the same training batch in an end-to-end backpropagation fashion.

The overall results are shown in Table~\ref{table:result1}. We can observe:
\textit{\textbf{(1)}} FDE achieves the worst recommendation performance and largest Params, where FDE is assigned the maximal embedding dimension 32 to all feature fields. This result demonstrates that allocating same dimension to all feature fields is not only memory inefficient, but introduces numerous noises into the model.
\textit{\textbf{(2)}} RaS, AutoDim-s, AutoDim performs better than MDE, DPQ, NIS, MGQE, AutoEmb. The major differences between these two groups of methods are: (i) the first group aims to assign different embedding dimensions to different feature fields, while embeddings in the same feature field share the same dimension; (ii) the second group attempts to assign different embedding sizes to different feature values within the same feature fields, which are based on the frequencies of feature values. The second group of methods surfer from several challenges: (ii-a) there are numerous unique values in each feature field, e.g., $2.7\times10^4$ values for each feature field on average in Criteo dataset. This leads to a huge search space (even after bucketing) in each feature field, which makes it difficult to find the optimal solution, while the search space for each feature field is $N=5$ in AutoDim; (ii-b) allocating dimensions solely based on feature frequencies (i.e., how many times a feature value appears in the training set) may lose other important characteristics of the feature; (ii-c) the feature values frequencies are usually dynamic and not pre-known in real-time recommender system, e.g., the cold-start users/items; and (ii-d) it is difficult to manage embeddings with different dimensions for the same feature filed.
\textit{\textbf{(3)}} AutoDim outperforms RaS and AutoDim-s, where AutoDim updates the architectural weights $\boldsymbol{\alpha}$ on the validation batch, which can enhance the generalization; AutoDim-s updates the $\boldsymbol{\alpha}$  with DLRS on the same training batch simultaneously, which may lead to overfitting;  RaS randomly search the dimensions, which has a large search space. AutoDim-s has much larger Params than AutoDim, which indicates that larger dimensions are more efficient in minimizing training loss.

To sum up, we can draw an answer to the first question: compared with the representative baselines, AutoDim achieves significantly better recommendation performance, and saves $70\%\sim80\%$ embedding parameters. These results prove the effectiveness of the AutoDim framework.

\begin{figure*}[t]
	\centering
	{\subfigure{\includegraphics[width=0.161\linewidth]{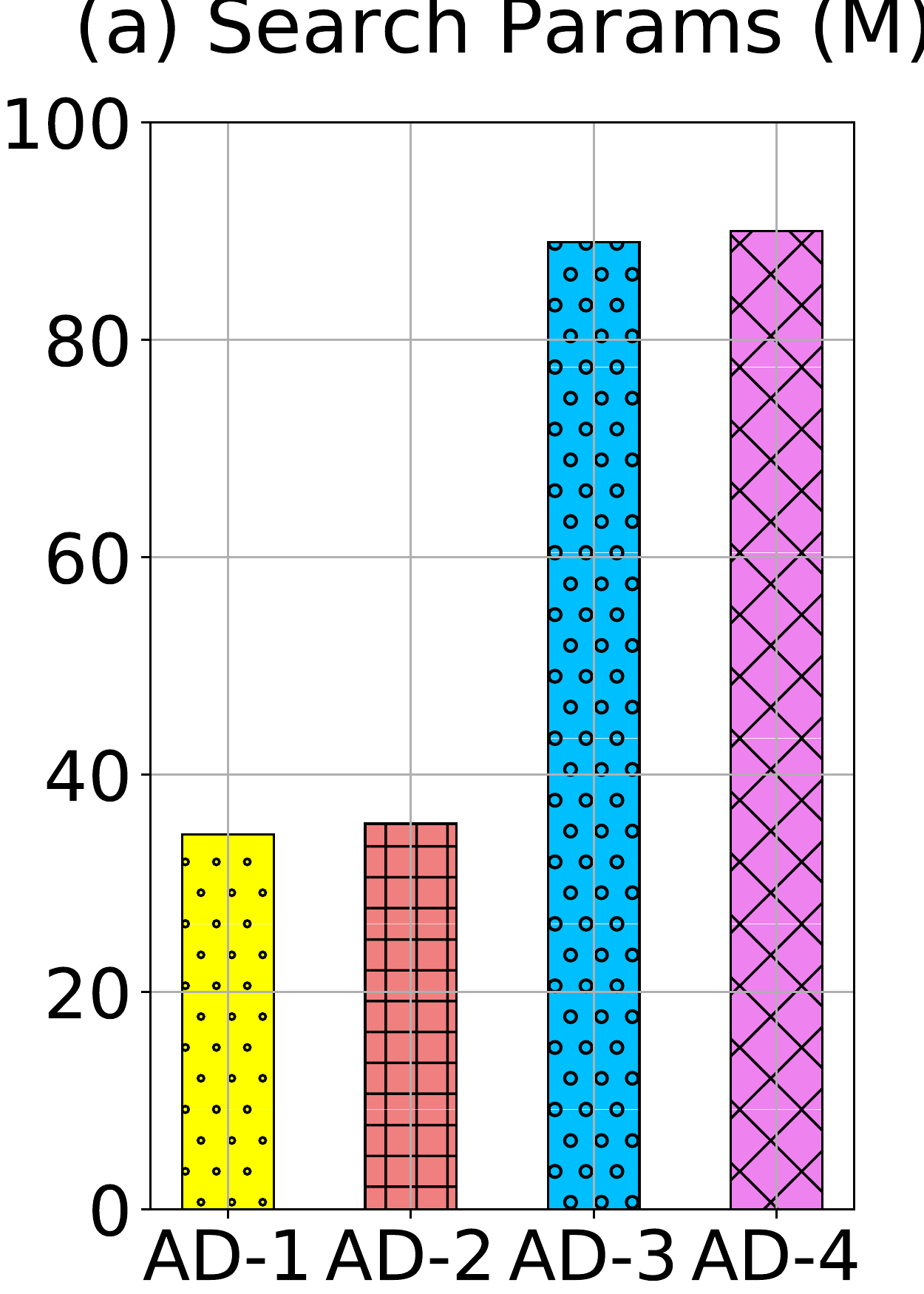}}}
	{\subfigure{\includegraphics[width=0.161\linewidth]{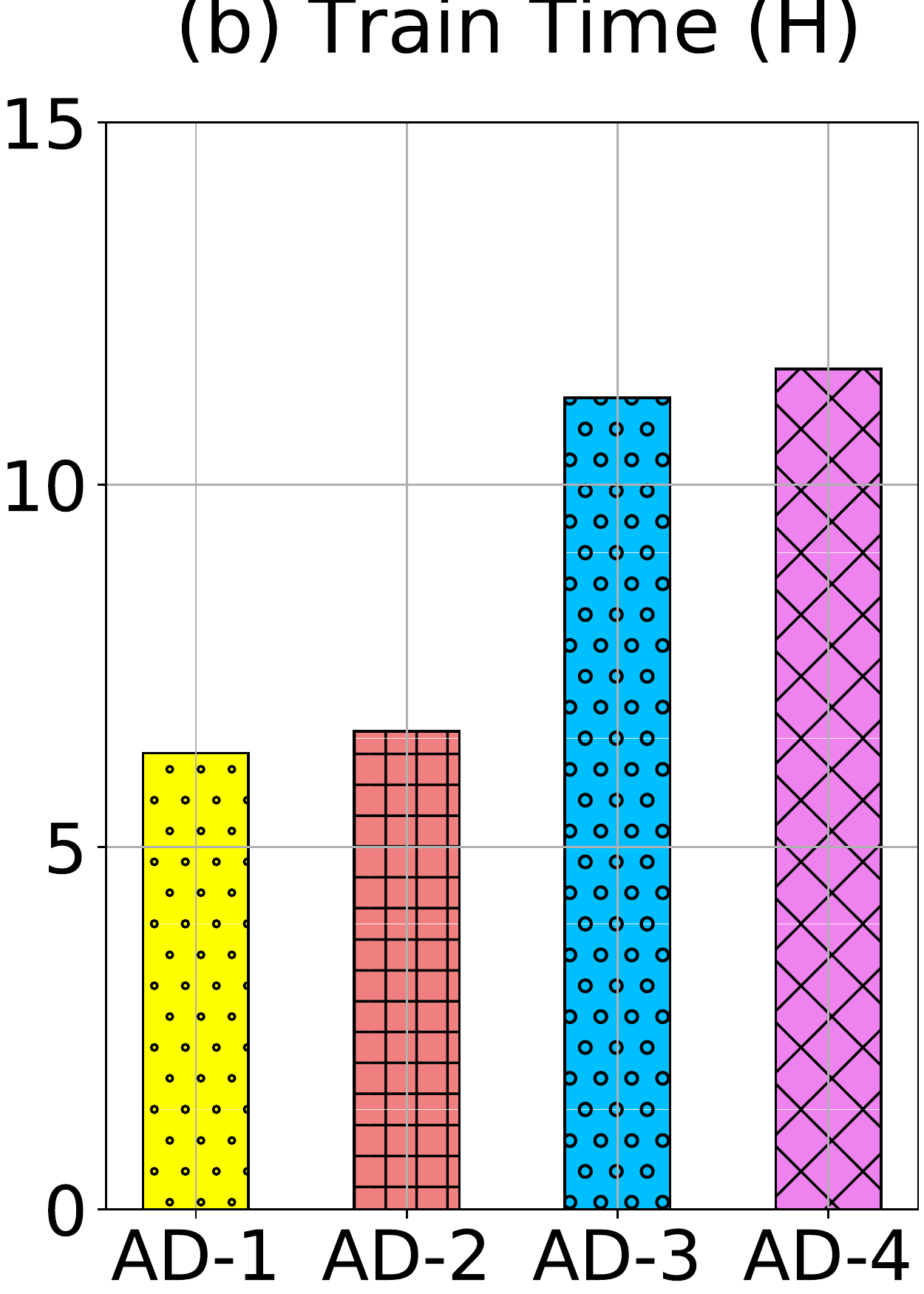}}}
	{\subfigure{\includegraphics[width=0.161\linewidth]{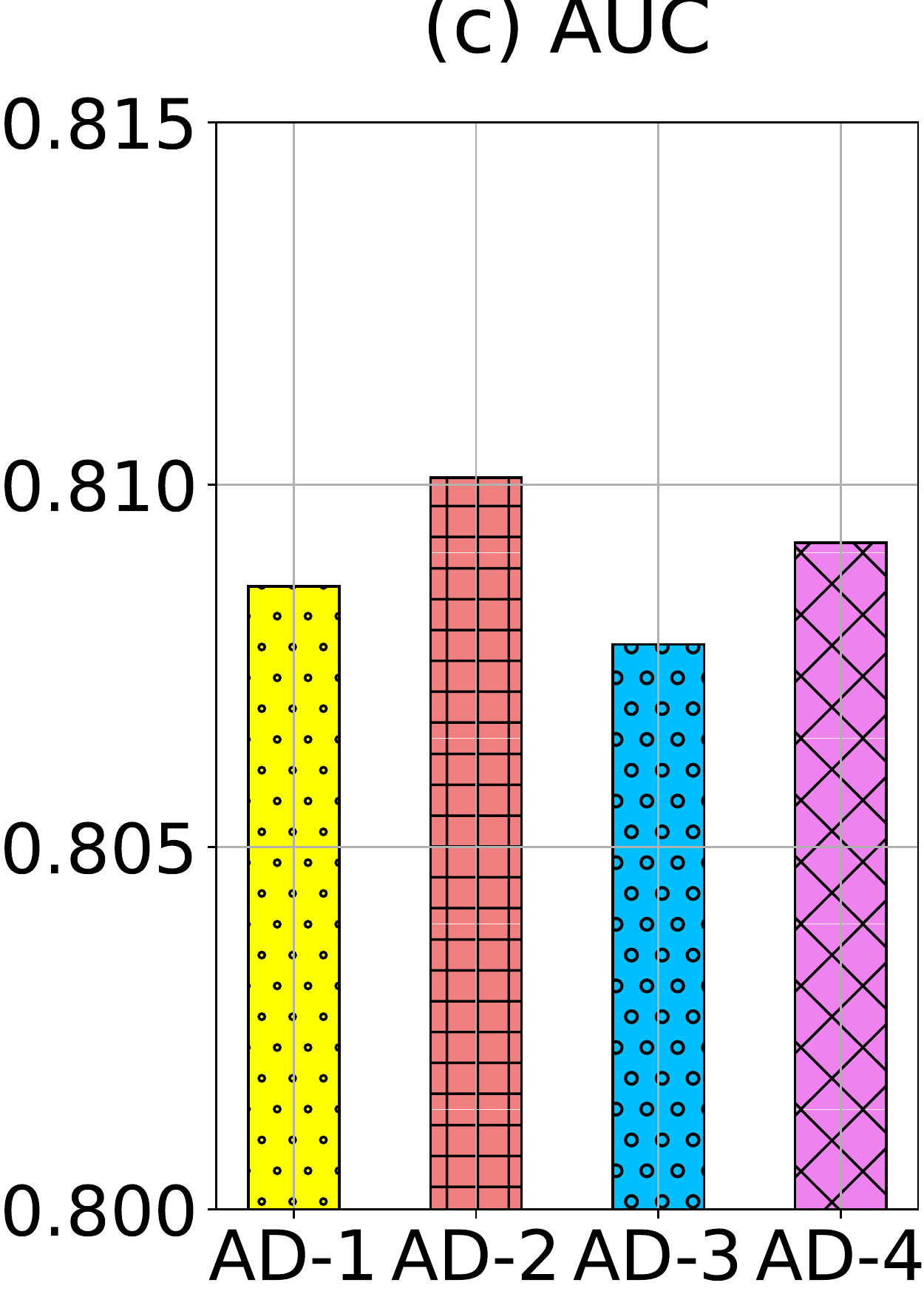}}}
	{\subfigure{\includegraphics[width=0.161\linewidth]{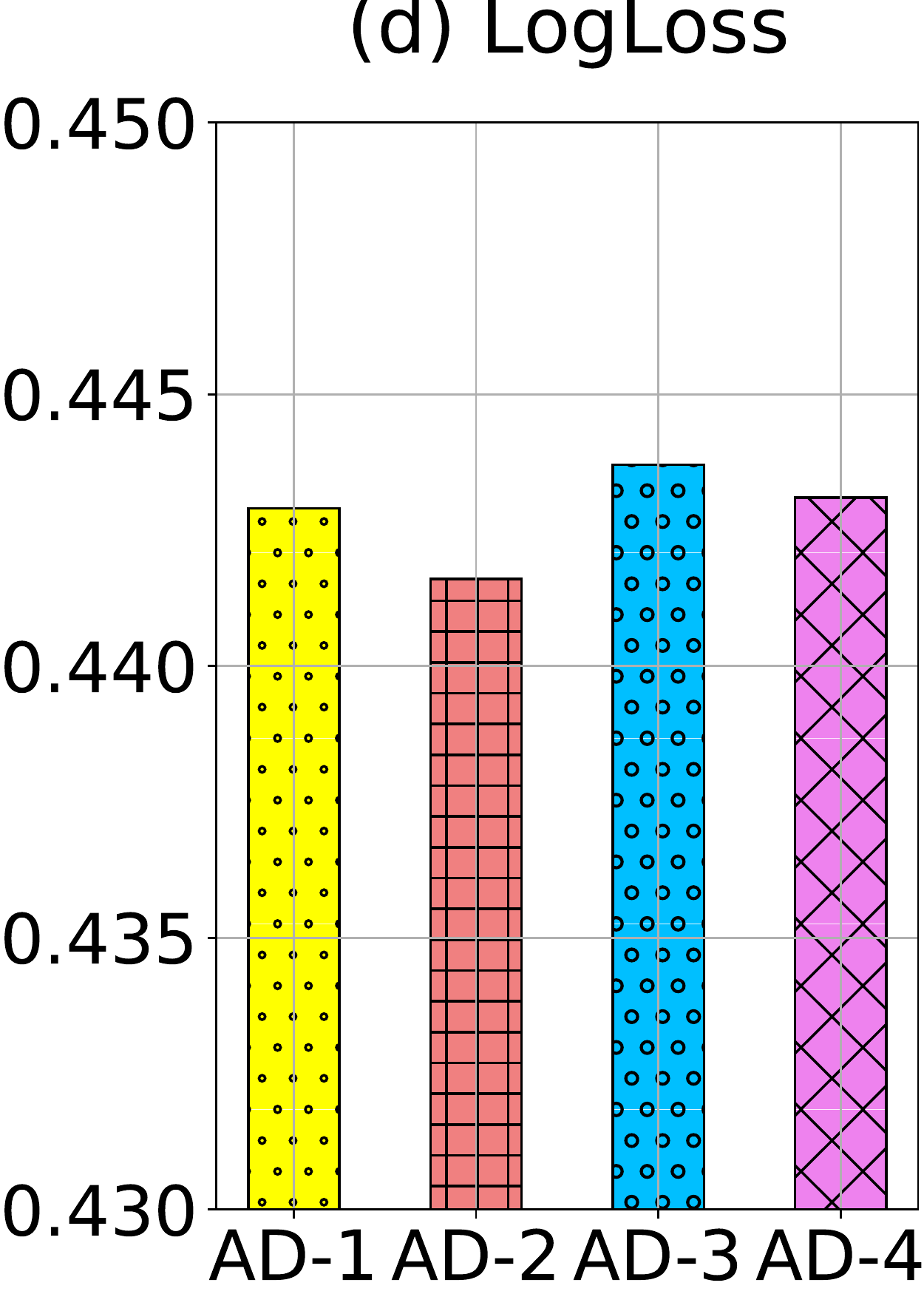}}}
	{\subfigure{\includegraphics[width=0.161\linewidth]{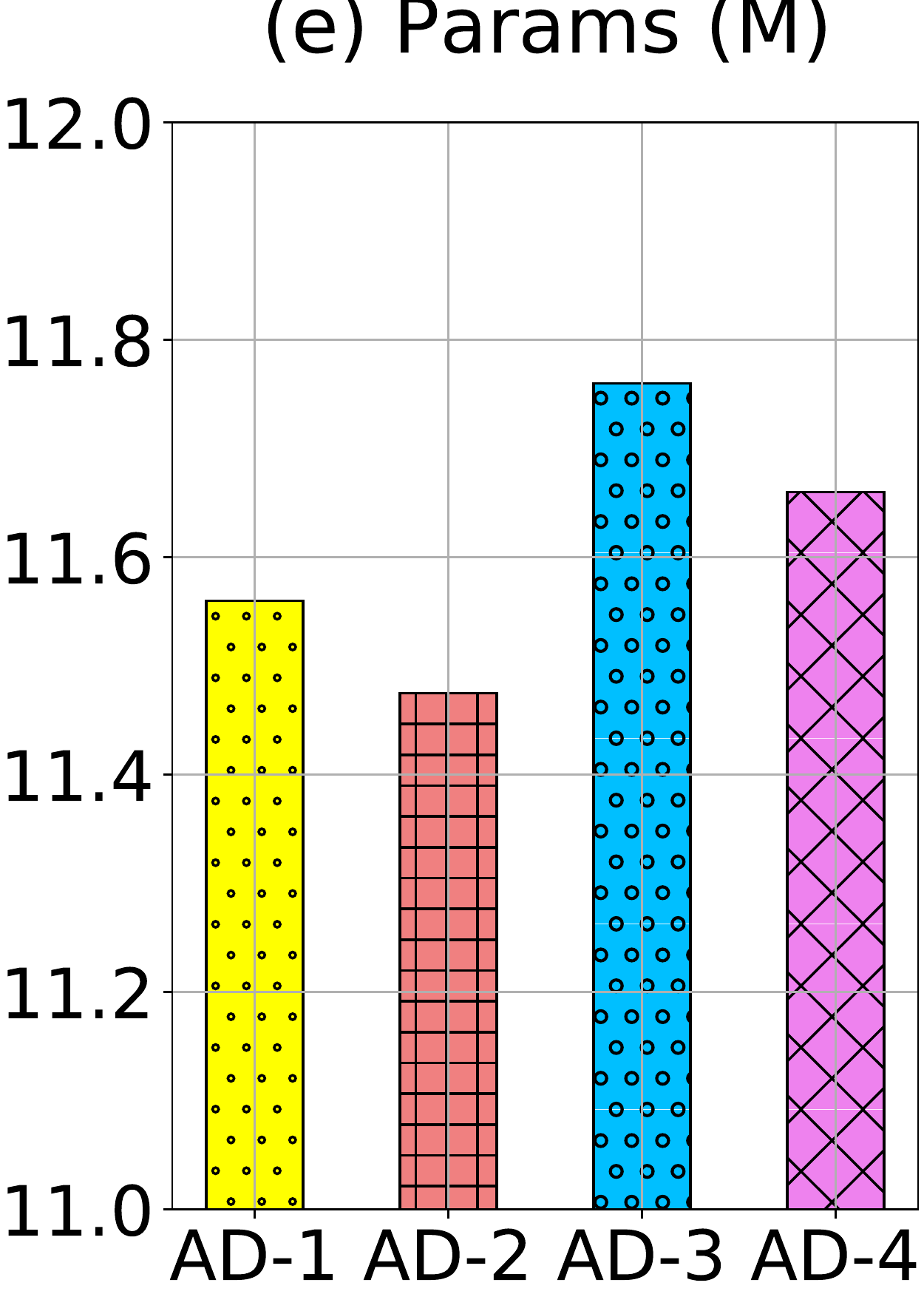}}}
	{\subfigure{\includegraphics[width=0.161\linewidth]{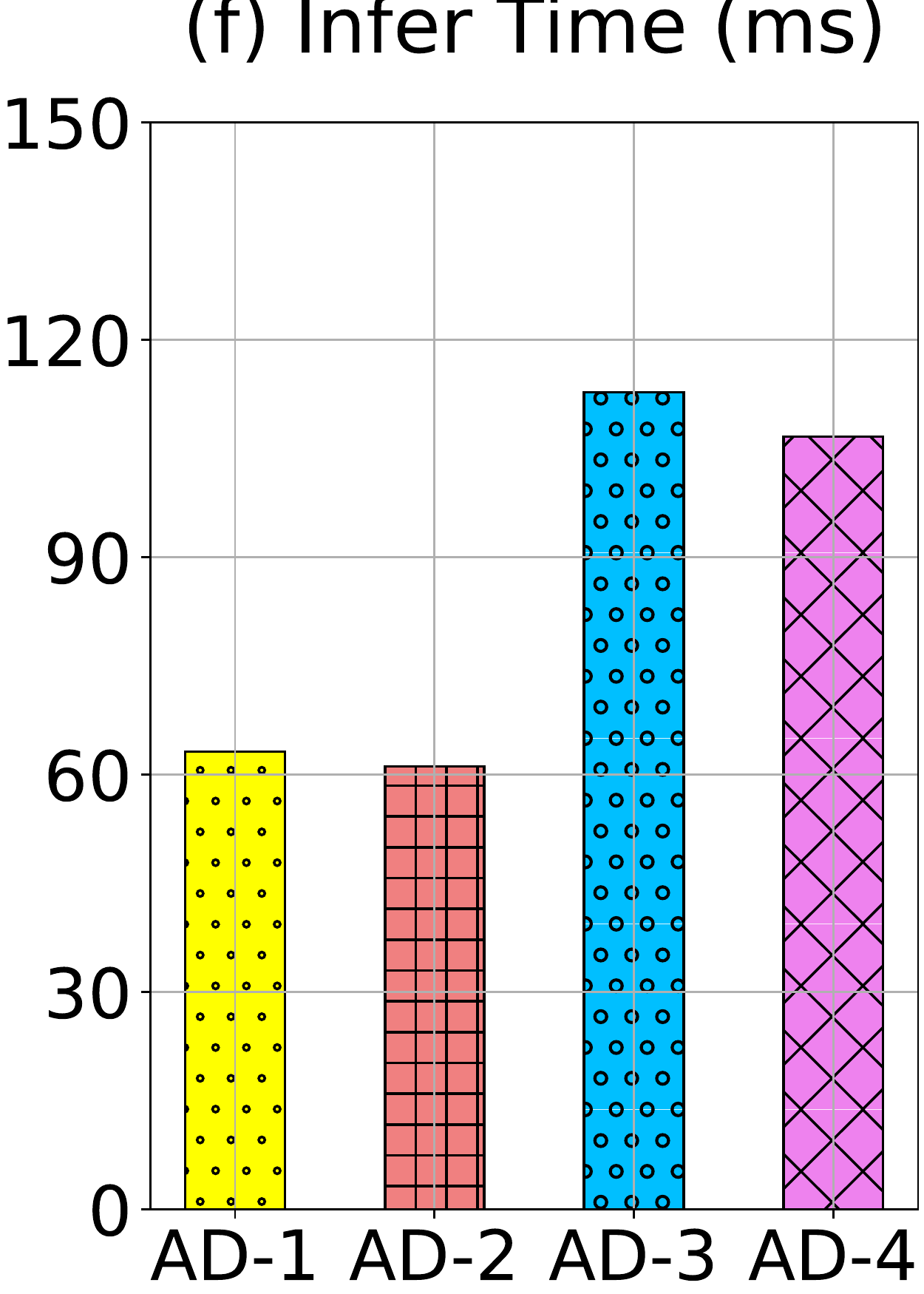}}}
	\caption{Component analysis of DeepFM on Criteo dataset. *(f) Infer time is averaged for one batch (batch size = 2,000)}\label{fig:Fig6_Component}
\end{figure*}

\subsection{Component Analysis (RQ2)}
\label{sec:RQ2}
In this paper, we propose two embedding lookup methods in Section~\ref{sec:Lookup} (i.e. separate embeddings v.s. weight-sharing embeddings) and two transformation methods  in Section~\ref{sec:Unifying} (i.e. linear transformation v.s. zero-padding transformation). In this section, we investigate their influence on performance. We systematically combine the corresponding model components by defining the following variants of AutoDim: 
(i) \textbf{AD-1}: weight-sharing embeddings and zero-padding transformation;
(ii) \textbf{AD-2}: weight-sharing embeddings and linear transformation;
(iii) \textbf{AD-3}: separate embeddings and zero-padding transformation;
(iv) \textbf{AD-4}: separate embeddings and linear transformation.

The results of DeepFM on the Criteo dataset are shown in Figure~\ref{fig:Fig6_Component}. We omit similar results on other models/datasets due to the limited space. We make the following observations: 
\textit{\textbf{(1)}} In Figure~\ref{fig:Fig6_Component} (a), we compare the embedding component parameters of in the \textit{dimension search stage}, i.e., all the candidate embeddings and the transformation neural networks shown in Figure \ref{fig:Fig3_Search1} or \ref{fig:Fig3_Search2}. We can observe that AD-1 and AD-2 save significant model parameters by introducing the weight-sharing embeddings, which also leads to a faster training speed in Figure~\ref{fig:Fig6_Component} (b). Therefore, weight-sharing embeddings can benefit real-world recommenders where exist thousands of feature fields and the computing resources are limited.
\textit{\textbf{(2)}} Compared with linear transformation, leveraging zero-padding transformation have slightly fewer parameters, and result in slightly faster training speed (e.g., AD-1 v.s. AD-2 in Figure~\ref{fig:Fig6_Component} (a) and (b)). However, we can observe the final DLRS architecture selected by AD-1 loses to that of AD-2 in Figure~\ref{fig:Fig6_Component} (c) AUC and (d) Logloss. The reason is that zero-padding embeddings may lose information when conduct inner product. For instance,  to compute the inner product of $a=[a_1,a_2,a_3]$ and $b=[b_1, b_2]$, we first pad $b$ to $b=[b_1, b_2, 0]$, then $<a,b>=a_1\cdot b_1 + a_2\cdot b_2 + a_3\cdot 0$, where the information $a_3$ is lost via multiplying 0. Models without element-wise product between embeddings, such as FNN~\cite{zhang2016deep}, do not suffer from this drawback.
\textit{\textbf{(3)}}  In Figure~\ref{fig:Fig6_Component} (e) and (f), we can observe that the final embedding dimensions selected by AD-2 save most model parameters and has the fastest inference speed.
\textit{\textbf{(4)}} From Figure~\ref{fig:Fig6_Component} (c) and (d), variants with weight-sharing embeddings have better performance than variants using separate embeddings. This is because the relatively front digits of its embedding space are more likely to be recalled and trained (as shown in Figure~\ref{fig:Fig2_Lookup} (b)), which enable the framework capture more essential information in these digits, and make optimal dimension assignment selection.

In summary, we can answer the second question: different component has its advantages, such as zero-padding has fastest training speed and uses least parameters, linear transformation has best performances on AUC/Logloss/Params metrics, and has good training speed and model parameters. If not specified, the results in other subsections are based on the AD-2, and we use linear transformation for RaS and AutoDim-s in Section~\ref{sec:RQ1}.

\begin{figure}[]
	\centering
	{\subfigure{\includegraphics[width=0.236\textwidth]{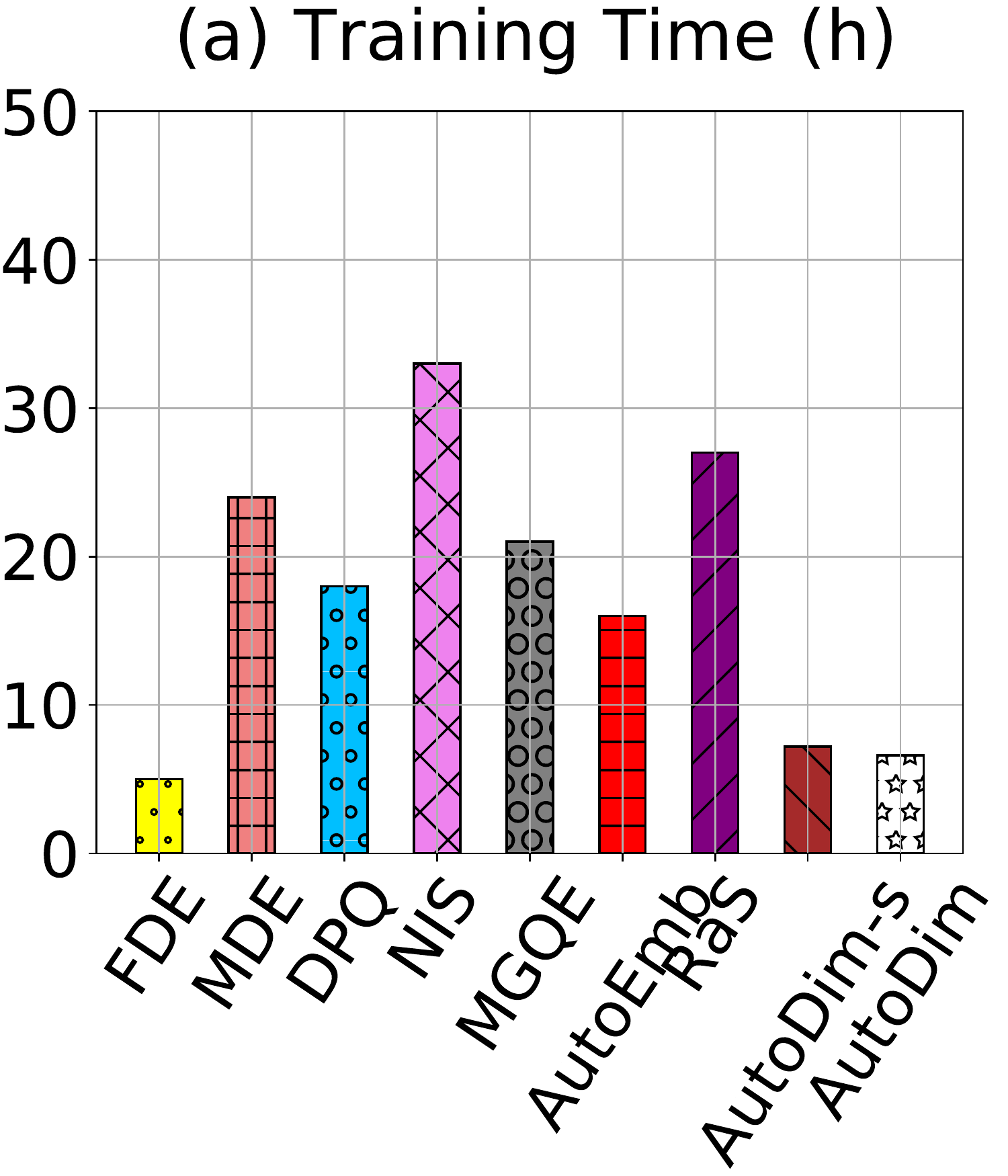}}}
	{\subfigure{\includegraphics[width=0.236\textwidth]{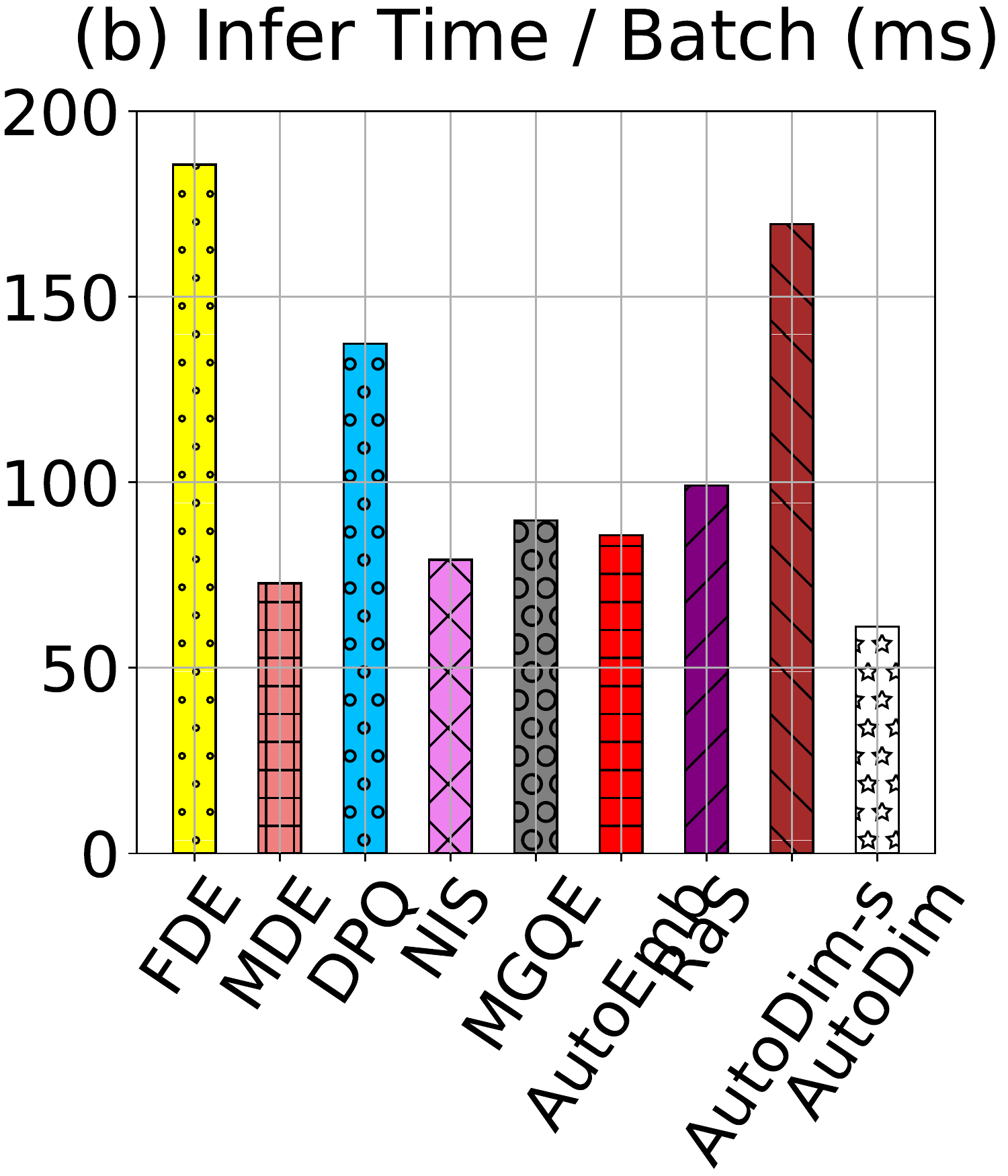}}}
	\caption{Efficiency analysis of DeepFM on Criteo dataset.} \label{fig:Fig5}
\end{figure}

\subsection{Efficiency Analysis (RQ3)}
\label{sec:Efficiency}
In addition to model effectiveness, the training and inference efficiency are also essential metrics for deploying a recommendation model into commercial recommender systems. In this section, we investigate the efficiency of applying search methods to DeepFM on Criteo dataset (on one Tesla K80 GPU). Similar results on other models/dataset are omitted due to the limited space. We illustrate the results in Figure~\ref{fig:Fig5}. 

For the training time in Figure ~\ref{fig:Fig5} (a), we can observe that AutoDim and AutoDim-s have fast training speed. As discussed in Section~\ref{sec:RQ1}, the reason is that they have a smaller search space than other baselines. FDE's training is fast since we directly set its embedding dimension as 32, i.e., no searching stage, while its recommendation performance is worst among all methods in Section~\ref{sec:RQ1}. For the inference time, which is more crucial when deploying a model in commercial recommender systems, AutoDim achieves the least inference time as shown in Figure~\ref{fig:Fig5} (b). This is because the final recommendation model selected by AutoDim has the least embedding parameters, i.e., the Params metric. 

To summarize, AutoDim can efficiently achieve better performance, which makes it easier to be launched in real-world recommender systems.

\begin{figure}[t]
	\centering
	{\subfigure{\includegraphics[width=0.236\textwidth]{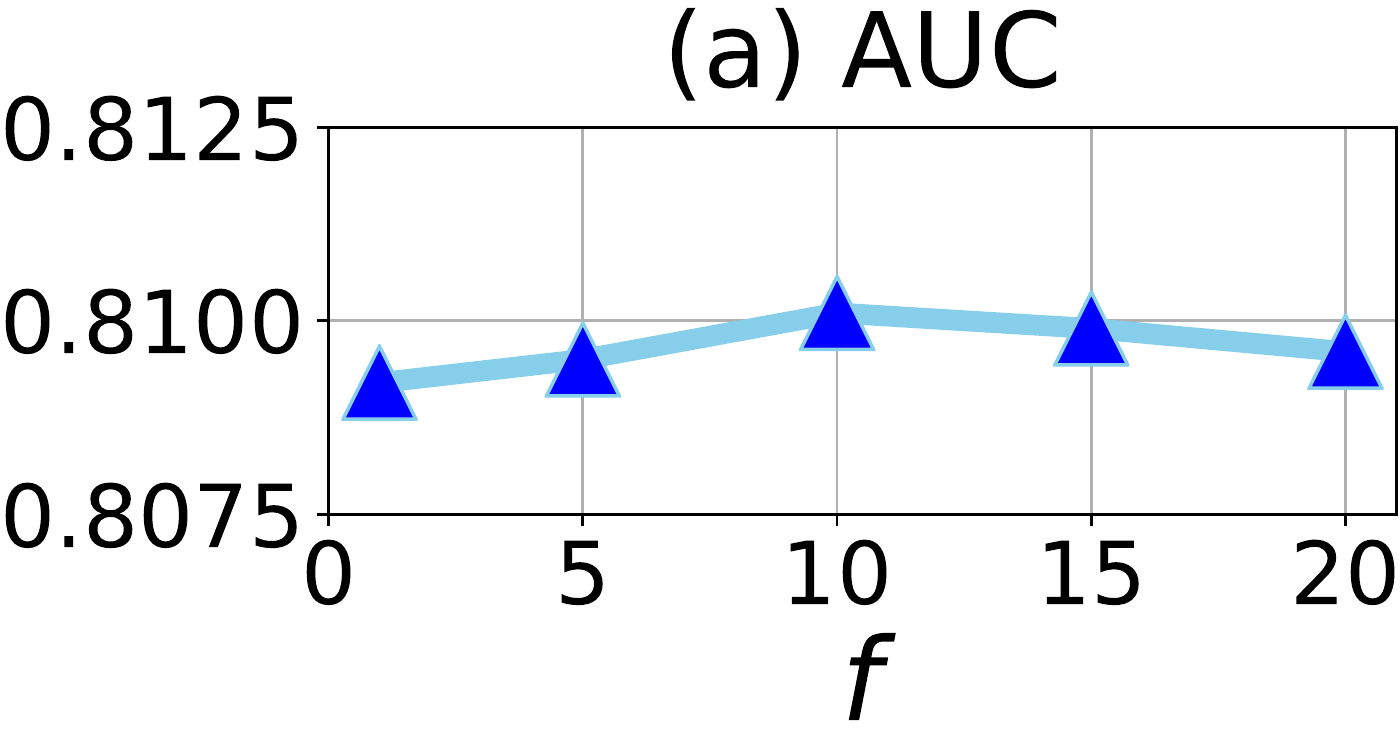}}}
	{\subfigure{\includegraphics[width=0.236\textwidth]{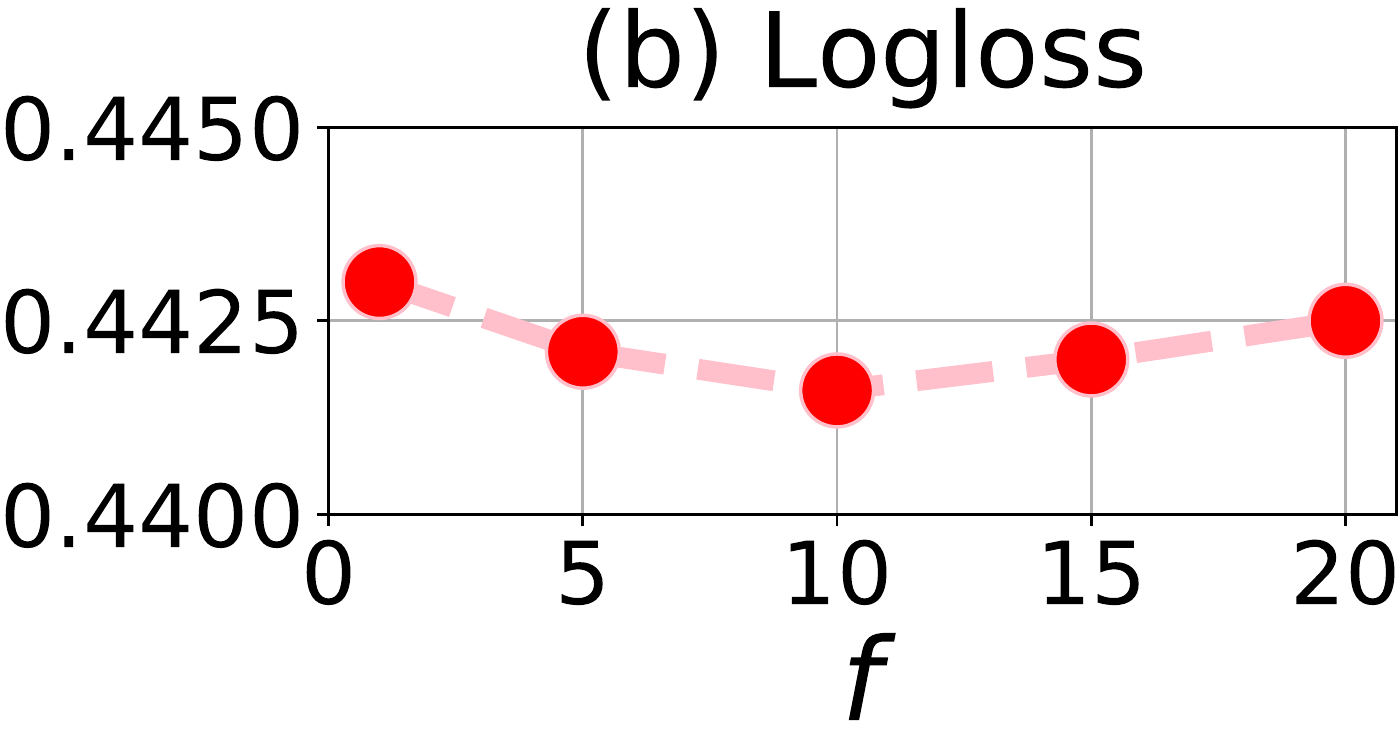}}}
	{\subfigure{\includegraphics[width=0.236\textwidth]{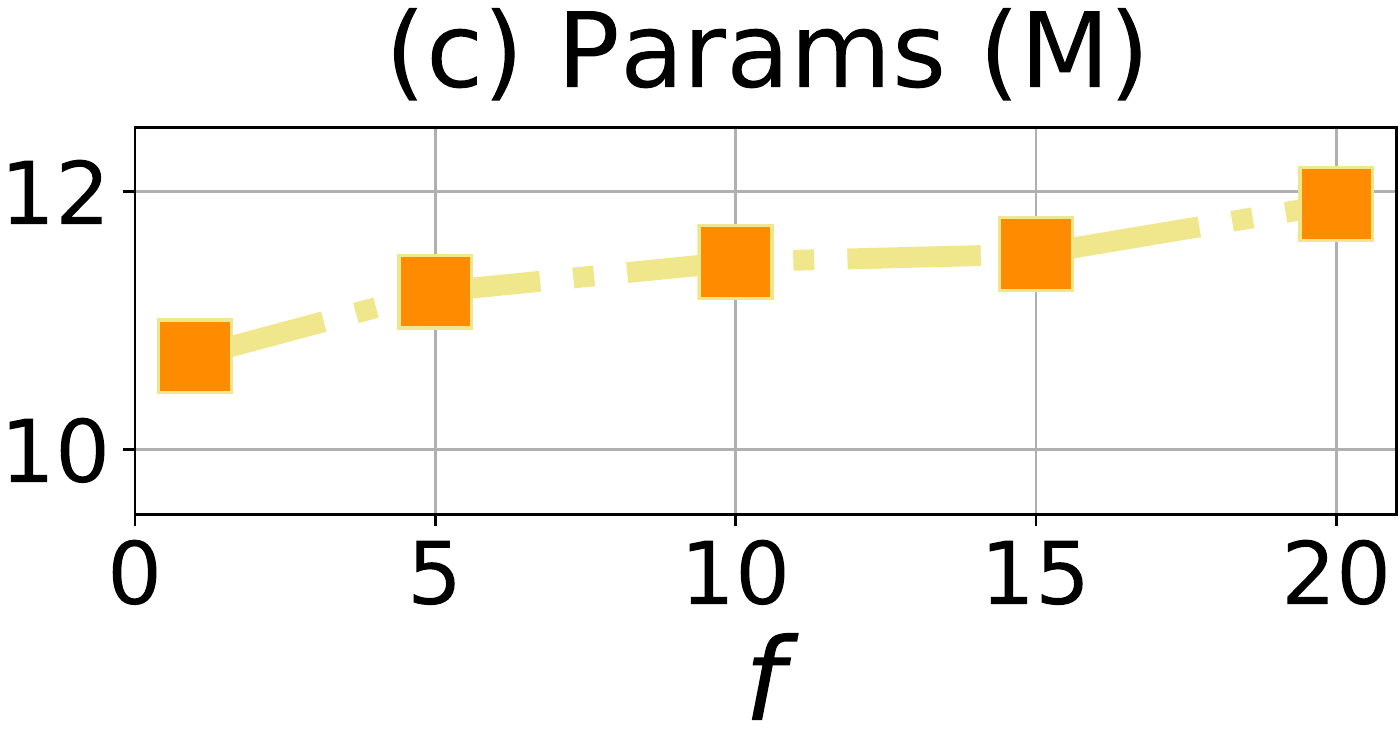}}}
	{\subfigure{\includegraphics[width=0.236\textwidth]{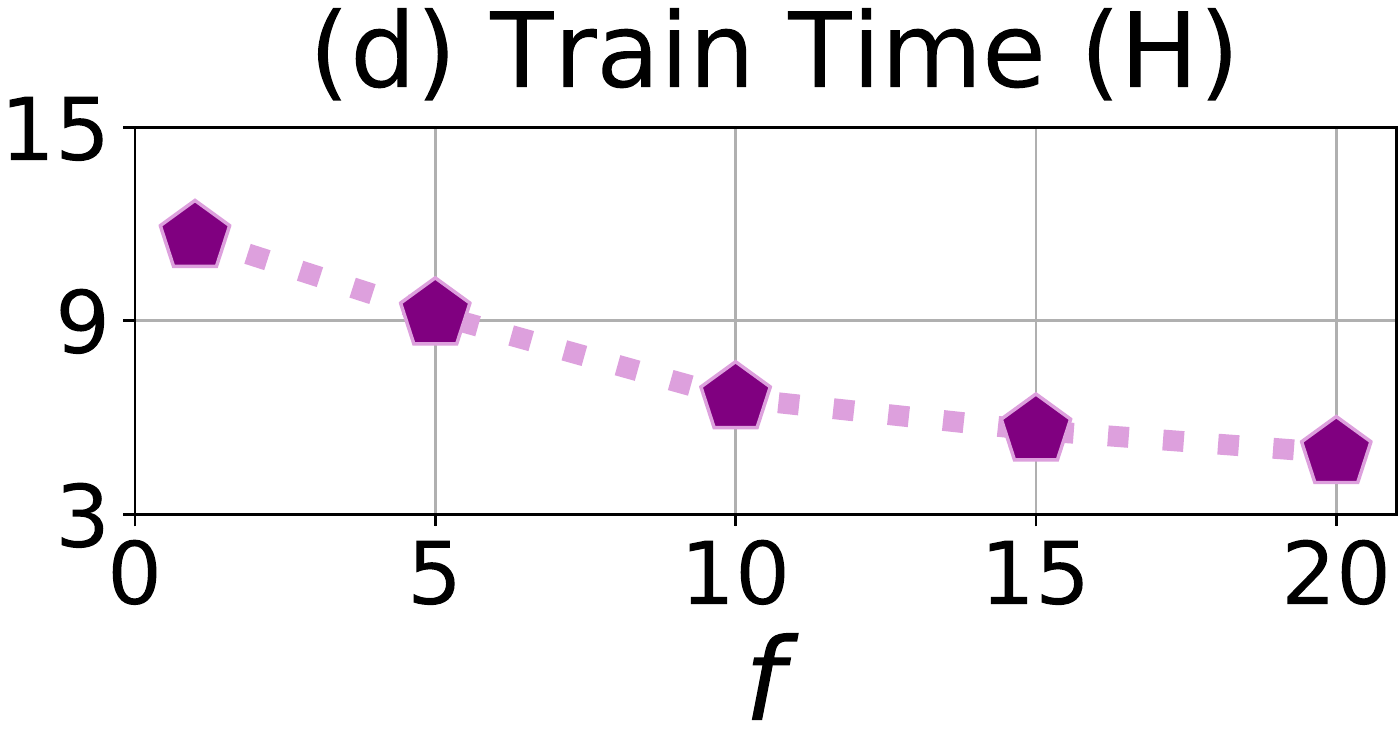}}}
	\caption{Parameter analysis of DeepFM on Criteo dataset.} \label{fig:Fig7_Parameter}
\end{figure}

\subsection{Parameter Analysis (RQ4)}
\label{sec:Parameter}
In this section, we investigate how the essential hyper-parameters influence model performance. Besides common hyper-parameters of deep recommender systems such as the number of hidden layers (we omit them due to limited space), our model has one particular hyper-parameter, i.e., the frequency to update architectural weights $\boldsymbol{\alpha}$, referred  to as $f$. In Algorithm~\ref{alg:DARTS}, we alternately update DLRS's parameters on the training data and update  $\boldsymbol{\alpha}$ on the validation data. In practice, we find that updating  $\boldsymbol{\alpha}$ can be less frequently than updating DLRS's parameters, which apparently reduces lots of computations, and also enhances the performance. 

To study the impact of $f$, we investigate how DeepFM with AutoDim performs on Criteo dataset with the changes of $f$, while fixing other parameters. Figure~\ref{fig:Fig7_Parameter} shows the parameter sensitivity results, where in $x$-axis, $f=i$ means updating $\boldsymbol{\alpha}$ once, then updating DLRS's parameters $i$ times. We can observe that the AutoDim achieves the optimal AUC/Logloss when $f=10$. In other words, updating  $\boldsymbol{\alpha}$ too frequently/infrequently results in suboptimal performance.  Figure~\ref{fig:Fig7_Parameter} (d) shows that setting $f=10$ can reduce $\sim50\%$ training time compared with setting $f=1$. 

Figure~\ref{fig:Fig7_Parameter} (c) shows that lower $f$ leads to lower Params, vice versa. The reason is that AutoDim updates  $\boldsymbol{\alpha}$ by minimizing validation loss, which improves the generalization of model~\cite{pham2018efficient,liu2018darts}. When updating $\boldsymbol{\alpha}$ frequently (e.g., $f=1$),  AutoDim tends to select smaller embedding size that has better generalization, while may has under-fitting problem; while when updating $\boldsymbol{\alpha}$ infrequently (e.g., $f=20$), AutoDim prefers larger embedding sizes that perform better on training set, but may lead to over-fitting problem. $f=10$ is a good trade-off between model performance on training and validation sets. Results of the other models/dataset are similar, we omit them because of the limited space.

\subsection{Transferability and Stability (RQ5)}
\subsubsection{Transferability of selected dimensions}
In this subsection, we investigate whether the embedding dimensions selected by a simple model (FM with AutoDim, say FM+AD) can be applied to the representative models, such as NFM~\cite{he2017neural}, PNN~\cite{qu2016product}, AutoInt~\cite{song2019autoint}, to enhance their recommendation performance. From Section~\ref{sec:RQ1}, we know the FM+AD can save $70\%\sim80\%$ embedding parameters.

The results are shown in Table \ref{tab:Transferability}, where "Model+AD" means assigning the embedding dimensions selected by FM+AD to this Model. We can observe that the performances of three models are significantly improved by applying embedding dimensions selected by FM+AD on two datasets. These observations demonstrate the transferability of embedding dimensions selected by FM+AD.

\subsubsection{Stability of selected dimensions}
To study whether the dimensions selected by AutoDim are stable, we run the \textit{search stage} of DeepFM+AutoDim on Criteo dataset with different random seeds. The Pearson correlation of selected dimensions from different seeds is around 0.85, which demonstrates the stability of the selected dimensions. 

\begin{table}[t]
	\caption{Transferability of selected dimensions. }
	\label{tab:Transferability}
	\begin{tabular}{@{}|c|cc|cc|@{}}
		\toprule[1pt]
		\multirow{2}{*}{Model} & \multicolumn{2}{c|}{Criteo} & \multicolumn{2}{c|}{Avazu} \\ \cmidrule(l){2-5} 
		& AUC & Logloss & AUC & Logloss \\ \midrule
		NFM & 0.8018 & 0.4491 & 0.7741 & 0.3846 \\
		NFM+AD & \textbf{0.8065*} & \textbf{0.4451*} & \textbf{0.7766*} & \textbf{0.3817*} \\ \midrule
		IPNN & 0.8085 & 0.4428 & 0.7855 & 0.3772 \\
		IPNN+AD & \textbf{0.8112*} & \textbf{0.4407*} & \textbf{0.7869*} & \textbf{0.3761*} \\ \midrule
		AutoInt & 0.8096 & 0.4418 & 0.786 & 0.3763 \\
		AutoInt+AD & \textbf{0.8116*} & \textbf{0.4403*} & \textbf{0.7875*} & \textbf{0.3756*} \\ \bottomrule[1pt]
	\end{tabular}
	\\``\textbf{{\Large *}}" indicates the statistically significant improvements (i.e., two-sided t-test with $p<0.05$).
\end{table}

\begin{table}[H]
	\caption{Embedding dimensions for Movielens-1m}
	\label{table:CaseStudy}
	\begin{tabular}{@{}|c|cc|c|@{}}
		\toprule[1pt]
		\multirow{2}{*}{feature   field} & \multicolumn{2}{c|}{W\&D (one field)} & AutoDim \\ \cmidrule(l){2-4} 
		& AUC & Logloss & Dimension \\ \midrule
		movieId & 0.7321 & 0.5947 & 8 \\
		year & 0.5763 & 0.6705 & 2 \\
		genres & 0.6312 & 0.6536 & 4 \\
		userId & 0.6857 & 0.6272 & 8 \\
		gender & 0.5079 & 0.6812 & 2 \\
		age & 0.5245 & 0.6805 & 2 \\
		occupation & 0.5264 & 0.6805 & 2 \\
		zip & 0.6524 & 0.6443 & 4 \\ \bottomrule[1pt]
	\end{tabular}
\end{table}

\subsection{Case Study (RQ6)}
\label{sec:CaseStudy}
In this section, we investigate whether AutoDim can assign larger embedding dimensions to more important features. Since feature fields are anonymous in Criteo and Avazu, we apply W\&D with AutoDim on MovieLens-1m dataset~\footnote{https://grouplens.org/datasets/movielens/1m/}. MovieLens-1m is a benchmark for evaluating recommendation algorithms, which contains users' ratings on movies. The dataset includes 6,040 users and 3,416 movies with 1 million user-item interactions. We binarize the ratings into a binary classification task, where ratings of 4 and 5 are viewed as positive and the rest as negative. There are $M=8$ categorical feature fields: movieId, year, genres, userId, gender, age, occupation, zip. Since MovieLens-1m is much smaller than Criteo and Avazu, we set the candidate embedding dimensions as $\{2,4,8,16\}$. 

To measure the contribution of a feature field to the final prediction, we build a W\&D model with only this field, train this model and evaluate it on the test set. A higher AUC and a lower Logloss means this feature field is more predictive for the final prediction. Then, we build a comprehensive W\&D model incorporating all feature fields, and apply AutoDim to select the dimensions for all feature fields. The results are shown in Table~\ref{table:CaseStudy}. It can be observed that: \textbf{\textit{(1)}} No feature fields are assigned 16-dimensional embedding space, which means candidate embedding dimensions $\{2,4,8,16\}$ are sufficient to cover all possible choices. \textit{\textbf{(2)}} Compared to the AUC/Logloss of W\&D with each feature field, we can find that AutoDim assigns larger embedding dimensions to important (highly predictive) feature fields, such as movieId and userId, vice versa. \textit{\textbf{(3)}} We build a full dimension embedding (FDE) version of W\&D, where all feature fields are assigned as the maximal dimension 16. Its performances are AUC=0.8077, Logloss=0.5383, while the performances of W\&D with AutoDim are  AUC=0.8113, Logloss=0.5242, and it saves 57\% embedding parameters.

In short, above observations validates that AutoDim can assign larger embedding dimensions to more predictive feature fields, which significantly enhances model performance and reduce embedding parameters.
\section{Related Work}
\label{sec:related_work}
In this section, we will discuss the related works. We summarize the works related to our research from two perspectives, say, deep recommender systems and AutoML for neural architecture search.

Deep recommender systems have drawn increasing attention from both the academia and the industry thanks to its great advantages over traditional methods~\cite{zhang2019deep}. Various types of deep learning approaches in recommendation are developed. Sedhain et al.~\cite{sedhain2015autorec} present an AutoEncoder based model named AutoRec. In their work, both item-based and user-based AutoRec are introduced. They are designed to capture the low-dimension feature embeddings of users and items, respectively. Hidasi et al.~\cite{hidasi2015session} introduce an RNN based recommender system named GRU4Rec. In session-based recommendation, the model captures the information from items' transition sequences for prediction. They also design a session-parallel mini-batches algorithm and a sampling method for output, which make the training process more efficient. Cheng et al.~\cite{cheng2016wide} introduce a Wide\&Deep framework for both regression and classification tasks. The framework consists of a wide part, which is a linear model implemented as one layer of a feed-forward neural network, and a deep part, which contains multiple perceptron layers to learn abstract and deep representations. Guo et al.~\cite{guo2017deepfm} propose the DeepFM model. It combines the factorization machine (FM) and MLP. The idea of it is to use the former to model the lower-order feature interactions while using the latter to learn the higher-order interactions. Wang et al.~\cite{wang2017your} attempt to utilize CNN to extract visual features to help POI (Point-of-Interest) recommendations. They build a PMF based framework that models the interactions between visual information and latent user/location factors. Chen et al.~\cite{chen2017attentive} introduce hierarchical attention mechanisms into recommendation models. They propose a collaborative filtering model with an item-level and a component-level attention mechanism. The item-level attention mechanism captures user representations by attending various items and the component-level one tries to figure out the most important features from auxiliary sources for each user. Wang et al.~\cite{wang2017irgan} propose a generative adversarial network (GAN) based information retrieval model, IRGAN, which is applied in the task of recommendation, and also web search and question answering.

The research of AutoML for neural architecture search can be traced back to NAS~\cite{zoph2016neural}, which first utilizes an RNN based controller to design neural networks and proposes a reinforcement learning algorithm to optimize the framework. After that, many endeavors are conducted on reducing the high training cost of NAS. Pham et al.~\cite{pham2018efficient} propose ENAS, where the controller learns to search a subgraph from a large computational graph to form an optimal neural network architecture. Brock et al.~\cite{brock2017smash} introduce a framework named SMASH, in which a hyper-network is developed to generate weights for sampled networks. DARTS~\cite{liu2018darts} and SNAS~\cite{xie2018snas} formulate the problem of network architecture search in a differentiable manner and solve it using gradient descent. Luo et al.~\cite{luo2018neural} investigate representing network architectures as embeddings. Then they design a predictor to take the architecture embedding as input to predict its performance. They utilize gradient-based optimization to find an optimal embedding and decode it back to the network architecture.
Some works raise another way of thinking, which is to limit the search space. The works~\cite{real2019regularized, zhong2018practical, liu2018progressive, cai2018path} focus on searching convolution cells, which are stacked repeatedly to form a convolutional neural network. Zoph et al.~\cite{zoph2018learning} propose a transfer learning framework called NASNet, which train convolution cells on smaller datasets and apply them on larger datasets. Tan et al.~\cite{tan2019mnasnet} introduce MNAS. They propose to search hierarchical convolution cell blocks in an independent manner, so that a deep network can be built based on them. Mixed Dimension Embedding~\cite{ginart2019mixed}, Differentiable Product Quantization~\cite{chen2019differentiable}, Neural Input Search~\cite{joglekar2019neural,cheng2020differentiable}, Multi-granular Quantized Embedding~\cite{kang2020learning}, and Automated Embedding Dimensionality Search~\cite{zhao2020autoemb} are designed for tuning the embedding layer of deep recommender system. But they aim to tune the embedding sizes within the same feature field, and we discuss the detailed differences and drawbacks of these models in \textit{Section~\ref{sec:RQ1}}.
\section{Conclusion}
\label{sec:conclusion}
In this paper, we propose a novel framework AutoDim, which targets at automatically assigning different embedding dimensions to different feature fields in a data-driven manner. In real-world recommender systems, due to the huge amounts of feature fields and the highly complex relationships among embedding dimensions, feature distributions and neural network architectures, it is difficult, if possible, to manually allocate different dimensions to different feature fields. Thus, we proposed an AutoML based framework to automatically select from different embedding dimensions. To be specific, we first provide an end-to-end differentiable model, which computes the weights over different dimensions for different feature fields simultaneously in a soft and continuous form, and we propose an AutoML-based optimization algorithm; then according to the maximal weights, we derive a discrete embedding architecture, and re-train the DLRS parameters. We evaluate the AutoDim framework with extensive experiments based on widely used benchmark datasets. The results show that our framework can maintain or achieve slightly better performance with much fewer embedding space demands.
\bibliographystyle{ACM-Reference-Format}
\bibliography{9Reference} 
\end{document}